\documentclass{ica-conf-2010}

\usepackage{natbib}
 \bibpunct{(}{)}{,}{a}{}{;}
\let\cite\citep

\usepackage{microtype}

\usepackage{subfig}
\usepackage[it]{subfigure}

\title{Vibrational and acoustical characteristics of the piano soundboard}
\author{%
  Kerem Ege and Xavier Boutillon}
\affiliations{Laboratory for the Mechanics of Solids UMR 7649, École polytechnique, 91128 Palaiseau Cedex, France\\\emph{kerem.ege@polytechnique.edu}}

\begin{document}
\twocolumn[{%
  \maketitle
  \begin{abstract}
The vibrations of the soundboard of an upright piano in playing condition are investigated.
It is first shown that the linear part of the response is at least 50 dB above its nonlinear component at normal levels of vibration. Given this essentially linear response, a modal identification is performed in the mid-frequency domain [300-2500]~Hz by means of a novel high resolution modal analysis technique~(Ege, Boutillon and David, JSV, 2009). The modal density of the spruce board varies between 0.05 and 0.01 modes/Hz and the mean loss factor is found to be approximately 2\%. 
Below 1.1 kHz, the modal density is very close to that of a homogeneous isotropic plate with clamped boundary conditions. Higher in frequency, the soundboard behaves as a set of waveguides defined by the ribs. A numerical determination of the modal shapes by a finite-element method confirms that the waves are localised between the ribs. The dispersion law in the plate above 1.1~kHz is derived from a simple waveguide model. We present how the acoustical coincidence scheme is modified in comparison with that of thin plates. The consequences in terms of radiation of the soundboard in the treble range of the instrument are also discussed.
  \end{abstract}
}]

\section{Introduction}
The purpose of this study is to describe the vibration regime of the soundboard of an
upright piano in playing condition in a large frequency range [300-2500]~Hz with only a few parameters. To this end, we have investigated the modal behaviour of the soundboard by means of a recently published high-resolution modal analysis technique~\cite{EGE2009}.\linebreak Compared to techniques based on the Fourier transform, it \linebreak avoids the customary frequency-resolution limitation and thus, gives access to a larger frequency-range and to a better precision on damping determinations. In the first section, we study the linearity of the board. Given the essential linear response, we present in the second section the results of two modal identifications of the soundboard from which we derive the modal density and the loss factor up to 2.5--3~kHz. The frequency evolution of the modal density of the piano soundboard reveals two well-separated vibratory regimes of the structure. The low-frequency behaviour (\emph{homogeneous isotropic plate}) is \linebreak presented in section~3 and the mid- and high-frequency behaviour (as exhibited by a \emph{set of waveguides}) in section~4.

\section{Linearity}
Nonlinear phenomena (such as jump phenomenon, hysteresis or internal resonance) appear when the vibration of a \linebreak bi-dimensional structure reaches amplitudes in the order of magnitude of its thickness \cite{TOU2002}. In the case of the piano, the soundboard displacement $w$ measured at the \linebreak bridge remains in a smaller range, even when played \emph{ff} and in the lower side of the keyboard. Askenfelt and Jansson report maximum values of displacement at the bridge $w_\text{max}\approx6\cdot 10^{-6}$~m in the frequency range [80-300]~Hz\linebreak \cite{ASK1992}. This maximum value is less than $10^{-3}$ times the board thickness. We can therefore assume that large displacements are far to be reached and the vibrations of the soundboard can be expected as linear to a high level of approximation.
\subsection{The technique}
In order to quantify experimentally the (non)linearity, we performed measurements on an upright piano soundboard. An \textit{exponential sine sweep} technique proposed by Farina \linebreak \cite{FAR2000}, mathematically proved by Rébillat~\emph{et~al.} \linebreak\cite{REB2010}, is used. It gives access both to the linear part of the impulse response of either system and to the nonlinear part of the response, that is the distortion level in the frequency-domain. 

The technique goes as follows: \\(a) Let's consider first a linear system excited by $x(t)$, a swept-sine of duration $T$ with initial and final angular frequencies $\omega_1$ and $\omega_2$: $x(t)=\sin{[\phi(t)]}$ with the instantaneous phase $\phi(t)=\omega_1 t + \cfrac{\omega_2-\omega_1}{T}\cfrac{t^2}{2}$\,. The impulse response $\gamma_\text{imp}(t)$ can be reconstructed by a deconvolution process: the measured signal $\gamma_\text{meas}(t)$ (acceleration for example) is convolved with the time-reversal of the excitation signal, that is $\gamma_\text{imp}(t)=\gamma_\text{meas}(t)\ast x(-t)$.\\(b) For a system with a weakly non-linear behaviour, Farina proposes to use a sine sweep for which the frequency varies exponentially with time -- \textit{exponential sine sweep} -- in order to separate the linear and nonlinear parts of the impulse response:
\begin{eqnarray}
\begin{aligned}
x(t)=\cos{[\phi(t)]}\\
\phi(t)=\cfrac{\omega_1 T}{\ln{(\omega_2/\omega_1)}}\,\left(e^{\frac{t}{T}\ln{(\omega_2/\omega_1)}}-1\right)-\pi/2
\end{aligned}
\end{eqnarray}
This signal verifies the fundamental property~\cite{REB2010}:
\begin{eqnarray}
\begin{aligned}
\forall k\in \mathbb{N}^*~,\qquad\cos{[k\phi(t)]}&=\cos{[\phi(t+\Delta\,t_k)]}\\
\mbox{where}\quad \Delta\,t_k&=\cfrac{T \ln{k}}{\ln{(\omega_2/\omega_1)}}
\end{aligned}
\end{eqnarray}
Multiplying the phase of a logarithmic sweep by a factor $k$ shifts it up in time by $\Delta\,t_k$. Rébillat~\emph{et al.} shown moreover that a logarithmic sweep to the power $n$, $x^n(t) = \cos^n[\phi(t)]$, can be written as a linear function of $\cos[k \phi(t)]$ with $k \in [1,n]$ \cite{REB2010}. This property is at the basis of the method for testing nonlinearity. The convolution product of the output signal $y(t)$ with the inverted excitation signal $x(-t)$ yields the linear impulse response preceded in time by the non-linear impulse responses of successive orders. If the excitation time $T$ is long enough, the responses do not overlap and can be separated in time by simple windowing. The experimental problem consists in separating the sources of nonlinearity.

\subsection{Results}
An upright piano of no particular merit has been put in a pseudo-anechoic room (anechoic walls and ceiling, ordinary ground). The piano was tuned normally but its strings were muted by strips of foam inserted between them or by woven in two or three places. Two configurations -- \{\emph{loudspeaker, room}\} and \{\emph{loudspeaker, piano, room}\} -- have been analysed with the following procedure. The electrical excitation of the loudspeaker was a logarithmic swept-sine [50-4000]~Hz with a 40 kHz sampling frequency and a $T=26~\text{s}$ duration. The amplitude of the loudspeaker was set at the beginning of the study in order to obtain displacements of the soundboard corresponding to the \textit{ff} playing: $\approx 10^{-6}~\text{m}$ at $\approx370$~Hz in this case. 

\begin{figure}[ht!]
      \centering \includegraphics[width=1\linewidth]{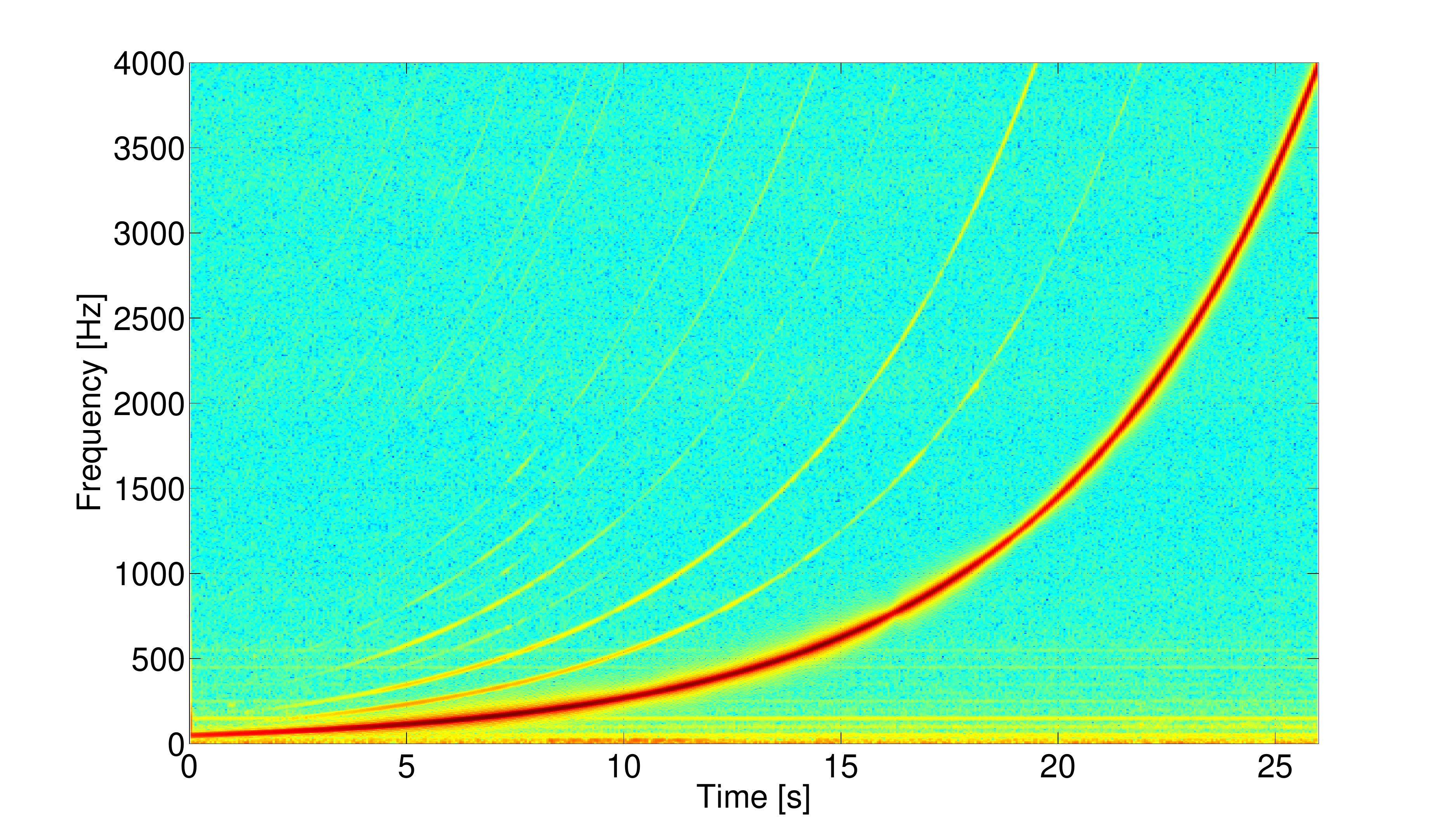}
      \caption[Distorsion]{Spectrogram of the acoustical response of the \emph{loudspeaker, room} system to an electrical signal. The distortion appears as successive harmonics of the swept-sine. The spectrogram is calculated with FFT over overlapping windows of 0.1~s with an 50\% overlap.}
      \label{fig:distorsion}
      \end{figure}

\begin{figure}[ht!]
      \centering\includegraphics[width=1\linewidth]{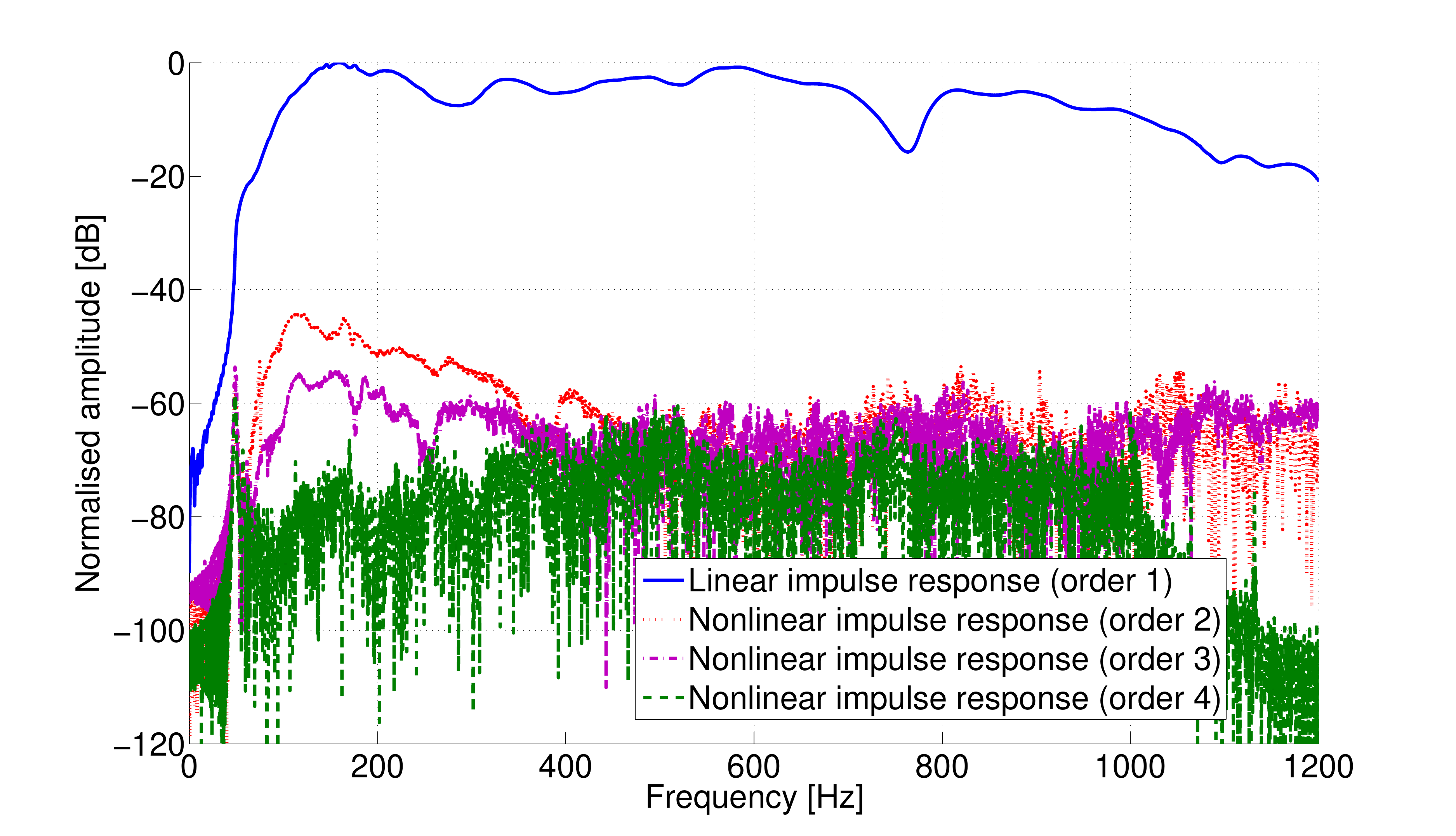}
      \caption[nonlinear_HP]{Spectra of the linear and the nonlinear responses shown in Fig.~\ref{fig:distorsion}. Except below 500~Hz, the nonlinear part of the response is $\approx$ 50-60~dB less than the linear part.}
      \label{fig:nonlinear_HP}
\end{figure}

In the first configuration \{\emph{loudspeaker, room}\}, the response of the room is measured with a microphone (prepolarised pressure-field 1/2'' -- Brüel \& Kj\ae r 4947) placed in front of the loudspeaker (Bose -- 802 Series II). The spectrogram of the complete pressure response of the room (without the piano) is shown in Fig.~\ref{fig:distorsion}. Some distortion is clearly visible which may safely be attributed to the loudspeaker rather than to the microphone. The spectra of the linear and nonlinear impulse responses separated with the method exposed above are shown in Fig.~\ref{fig:nonlinear_HP}. By convention, the non-linear response of order $n$ as displayed at frequency $f$ in these spectra, is the response to a sinusoid at $f$, measured at frequency $nf$ in a Fourier transform of the response. In other words, what is common to points belonging to different curves with the same abscissa is the frequency of the excitation signal. Except below 500~Hz where the distortion of the loudspeaker is large, the nonlinear response level is about 50-60~dB lower than the linear contribution.

\begin{figure}[!ht]
\centering
  \includegraphics[width=1\linewidth]{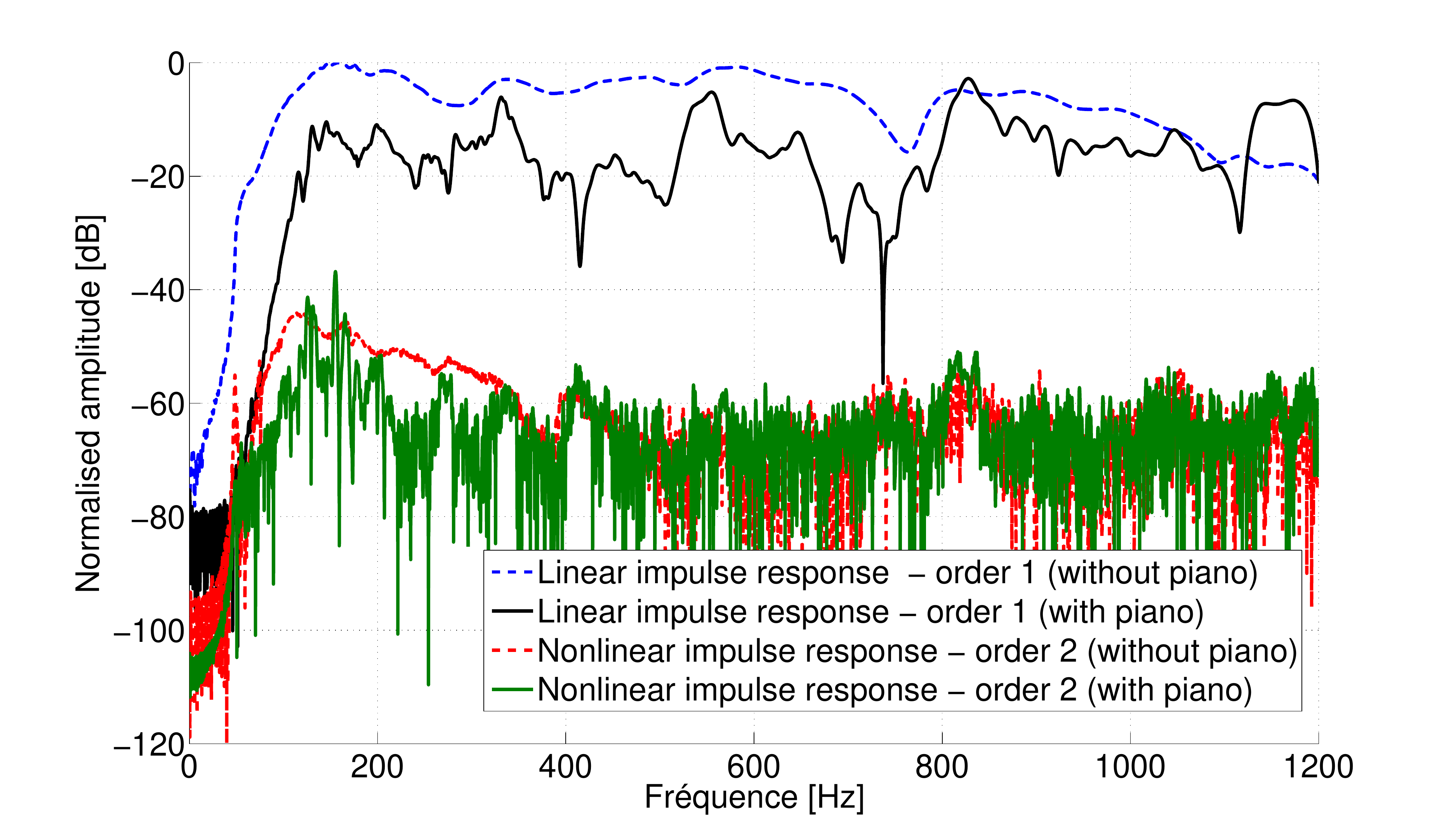}
  \caption[nonlinear_total]{Nonlinearities of the two systems \{\emph{loudspeaker, room}\} and \{\emph{loudspeaker, piano, room}\}. The distorsion rate is comparable in both situations.}
  \label{fig:nonlinear_total}
\end{figure}

In the second configuration \{\emph{loudspeaker, piano, room}\}, the motion of the soundboard was measured with an accelerometer (Brüel \& Kj\ae r 4393) put mid-way between two adjacent ribs, at $\approx10$~cm from the bridge, close to the F\#4 strings (fundamental frequency of $\approx370$~Hz). The spectra of the linear and nonlinear contributions to the response in each configuration (with and without the piano) are shown in Fig.~\ref{fig:nonlinear_total}. The distortion level appears to be approximately the same in both configurations. This shows that the soundboard intrinsic nonlinearity (distrosion rate) is of the order of -60 dB.

To conclude this preliminary study, it appears that a linear model is sufficient to predict the vibratory behaviour of a piano soundboard in playing situations, within the precision of customary measurements. Given this essentially linear character of the response, modal identifications of the soundboard have been performed. When a loudspeaker was used to excite the piano, the excitation level was comparable to the one used in this linearity study and the linear contribution was extracted. The ordinary impulse excitation does not permit to separate the linear and nonlinear contributions. Nevertheless, in the light of the results presented above, and considering the small amplitudes of displacement caused by the impacts on the tables (typically less than $8\cdot10^{-6}~\text{m}$, mostly due to a very low-frequency displacement and still less than 1/100 of the board thickness) we consider the linear approximation is also verified.

\section{Modal identification by a high-resolution method}
\subsection{The method}
\begin{figure}[!ht]
\centering
  \includegraphics[width=1\linewidth]{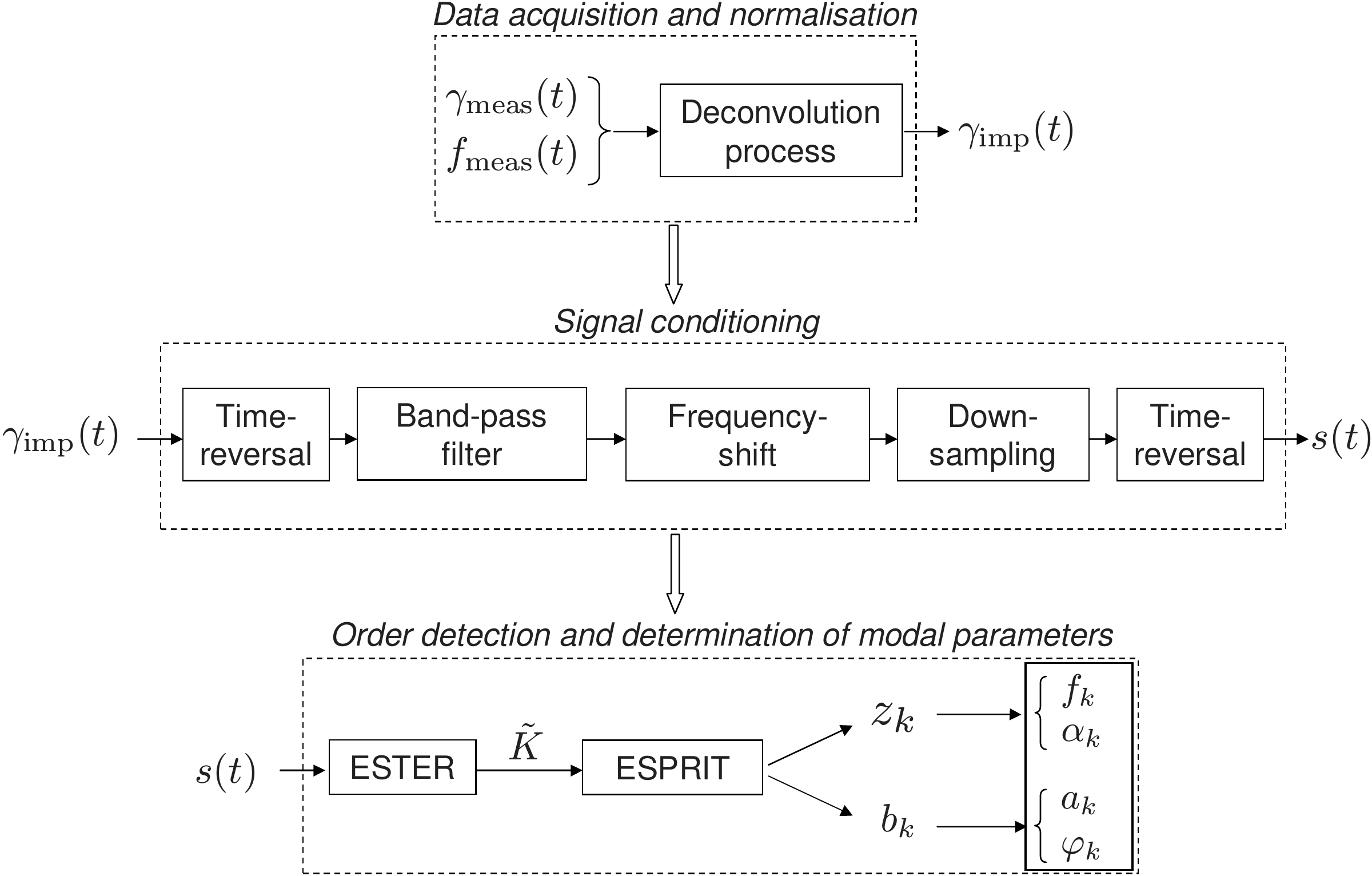}
  \caption[blockdiagram]{Block diagram of the high-resolution modal analysis method (after~\cite{EGE2009})}
  \label{fig:blockdiagram}
\end{figure}

The modal behaviour of the upright soundboard is investigated by means of a recently published high-resolution modal analysis technique~\cite{EGE2009} which avoids the frequency-resolution limitations of the Fourier transform. This new technique is particularly well suited for structures made of moderately damped materials such as spruce, and at frequencies where the modal overlap is high (more than 30\%). Based on the ESPRIT algorithm \cite{ROY1989}, it assumes that the signal is a sum of complex exponentials and white noise; it projects the signal onto two subspaces: the subspace spanned by the sinusoids (signal subspace) and its supplementary (noise subspace). Rotational invariance property of the signal subspace is used to estimate the modal parameters (frequencies, damping factors and complex amplitudes). The dimensions of both subspaces must be chosen~\emph{a priori} and the quality of the estimation depends on a proper choice for these parameters. The best choice for the dimension of the modal subspace is the number of complex exponentials actually present in the signal. This number, called $\tilde{K}$, is twice the number of decaying sinusoids. It is therefore advisable to estimate this number prior to the analysis. This is done by means of the ESTER technique~\cite{BAD2006}, recently developed: it consists in minimising the error on the rotational invariance property of the signal subspace spanned by the sinusoids. The block diagram of the method given in Figure~\ref{fig:blockdiagram} describes the three main steps of the method: (a) reconstruction of the acceleration impulse response (b) signal conditioning (c) order detection and determination of modal parameters. 

\subsection{Impulse excitation}\label{sec:impulse_excit}
\begin{figure}[!ht]
\centering
  \includegraphics[width=0.8\linewidth]{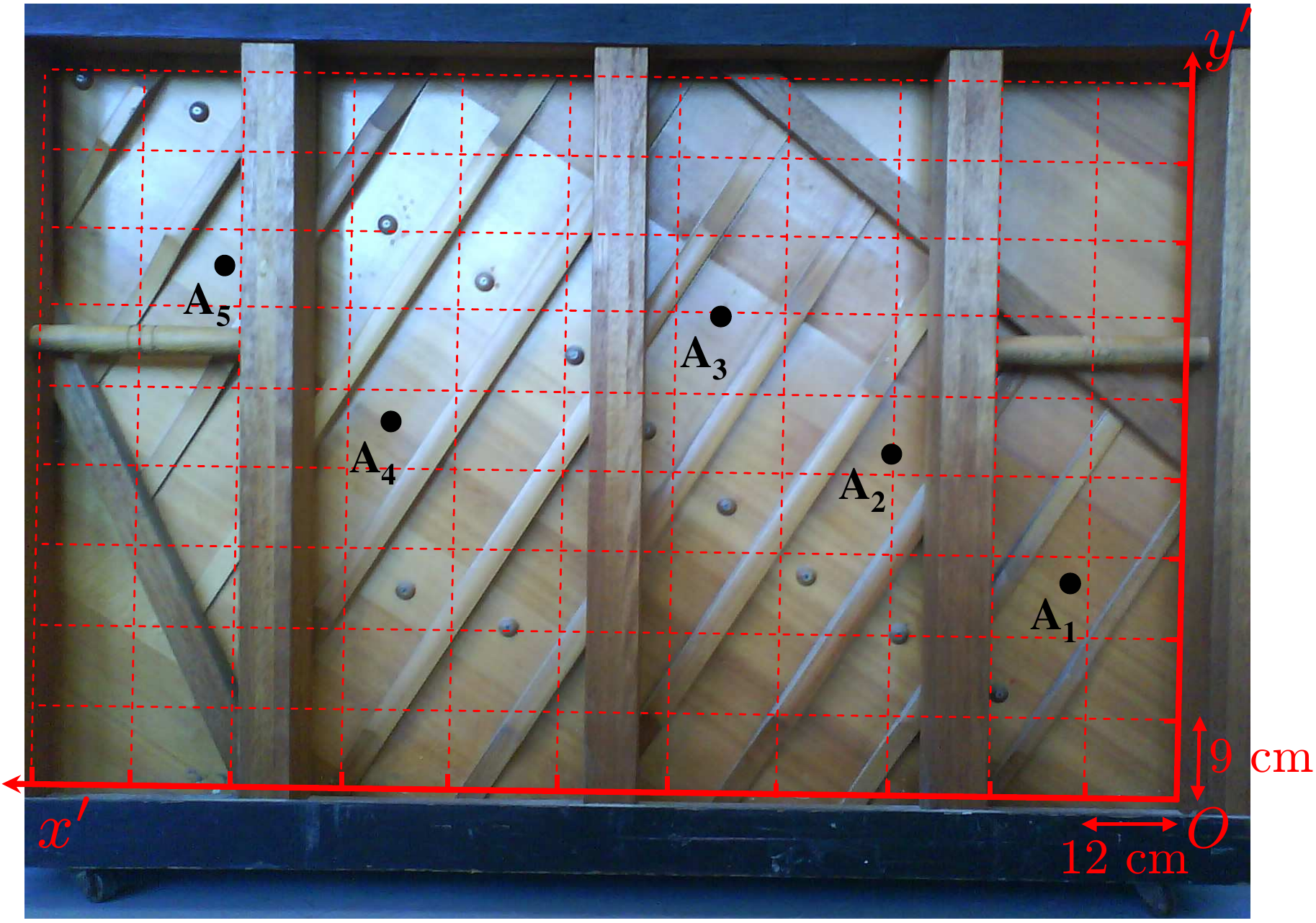}
  \caption{Rear view of the upright piano, with the mesh for modal analysis (in red) and the locations of the five accelerometers (in black).}
  \label{fig:table_exp_maillage}
\end{figure}

The experimental study presented here aims at estimating the modal parameters (modal frequencies, modal dampings and modal shapes) of the upright piano soundboard in the [0-500]~Hz frequency range. The piano was put in a pseudo-anechoic room and excited with an impact hammer (Kistler -- type 9722A) at the nodes of a rectangular mesh of $12\times10$ points regularly spaced (Figure~\ref{fig:table_exp_maillage}). The motion of the board is measured at five points with accelerometers ({two B\&K~2250A-10 and three B\&K~4393}) located in different zones of the board (Figure~\ref{fig:table_exp_maillage}).

\begin{figure}[!ht]
\centering
  \includegraphics[width=1\linewidth]{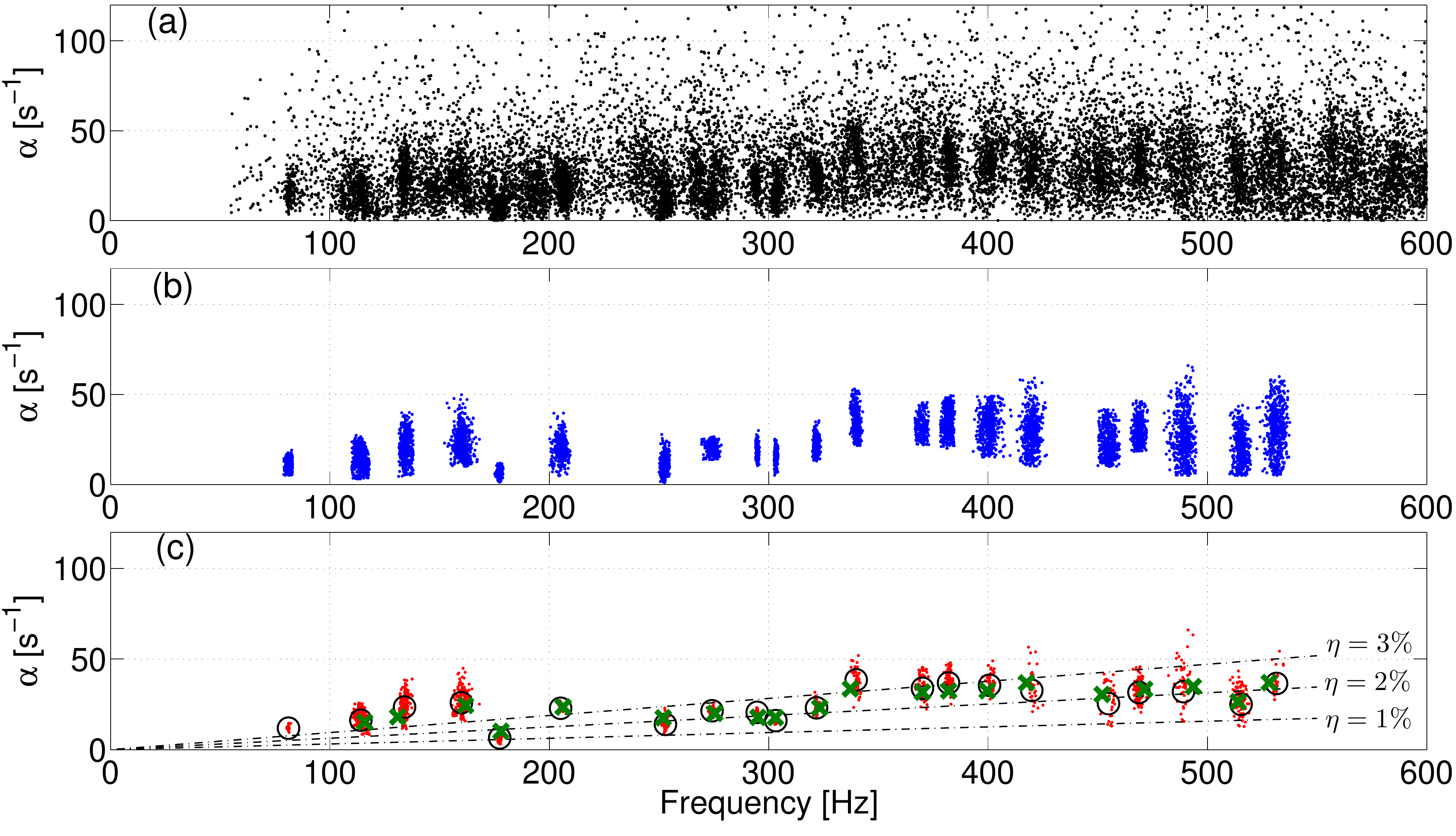}
  \caption{High-resolution modal analysis for the [0-600]~Hz frequency-band~(\emph{impulse excitation}). Modal frequencies/damping factors map. (a)~first \emph{rough} analysis. (b) After narrow band-pass filtering. (c)~After suppression of the (low-precision) estimations in nodal regions. $\circ$~: retained modal parameters. {\color[rgb]{0,0.5,0}$\times$}~: weighted mean of the modal parameters estimated at four points of the soundboard (\emph{acoustical excitation}, see following section). --\,$\cdot$\,--$\,$: constant loss factors ($\eta=$~1 to 3~\%).} 
  \label{fig:res_esprit_piano}
\end{figure}

\begin{figure}[!ht]
  \includegraphics[width=1\linewidth]{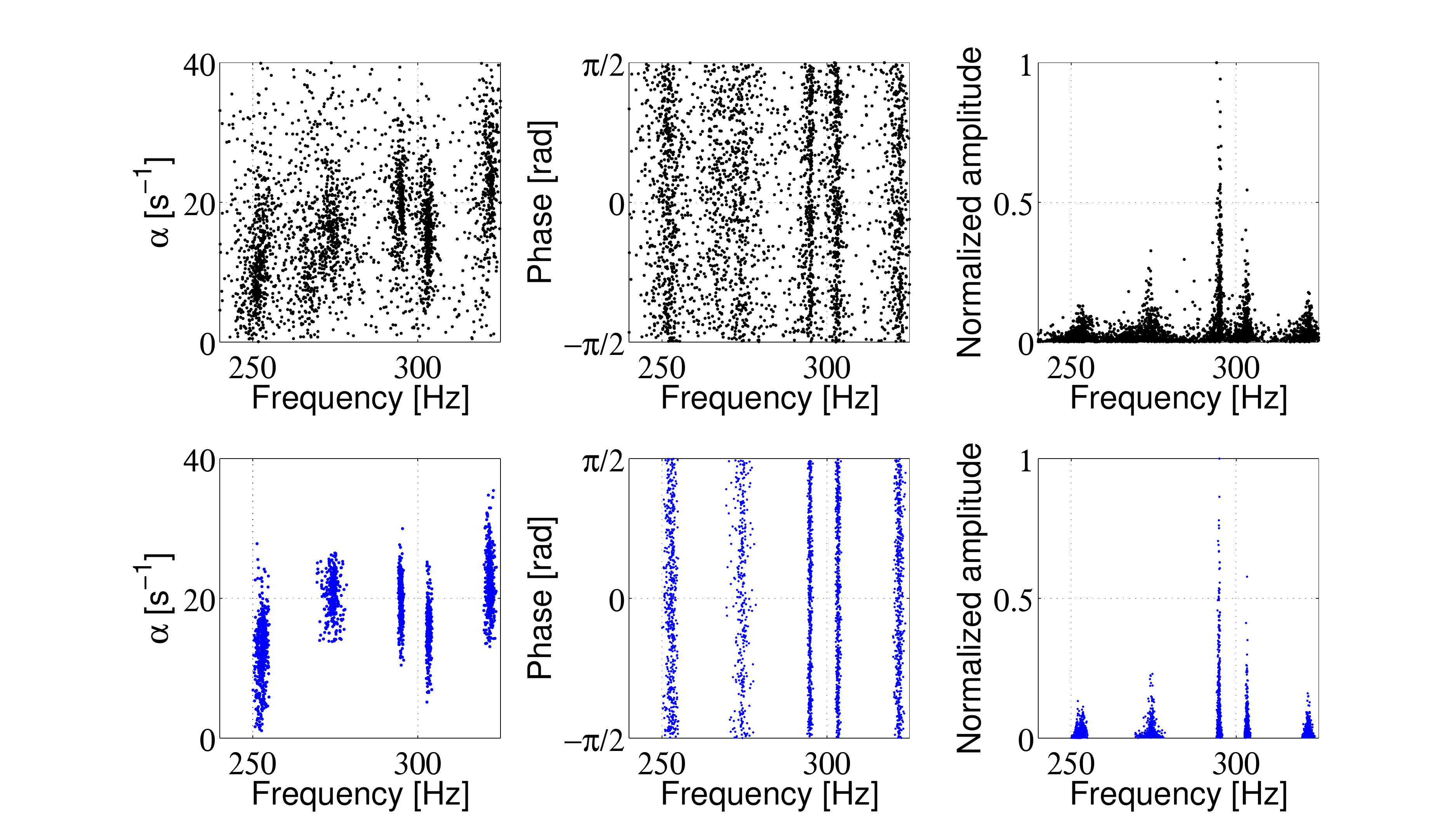}
  \caption{Necessity of the narrow band-pass filtering step. Results for five modes in the~[230-330]~Hz frequency-band before filtering (top diagrams), after filtering (bottom diagrams).}
    \label{fig:esprit_piano_bandeetroite_2}
\end{figure}

For each of the $120\times5$ measures, the impulse response is reconstructed and analysed with the high-resolution modal analysis method. Results are summarised in Figure~\ref{fig:res_esprit_piano}. In order to measure the damping with some precision, it proved necessary to band-filter the impulse responses prior to analysis, as shown by the comparison between (a) and (b) of Fig.~\ref{fig:res_esprit_piano} and as illustrated by Fig.~\ref{fig:esprit_piano_bandeetroite_2}.

Except for the first four low-frequency resonances -- for which the rim probably adds non-negligible losses to the distributed ones -- modal dampings are very close of the loss factors of the spruce (about 1-3\%). The average modal spacing (inverse of the modal density) is about 22~Hz for these $21$ lowest modes, in agreement with comparable low-frequency studies ($\approx 24.8$~Hz for a similar upright piano \cite{DER1997}, $\approx 22.3$~Hz for a baby grand one~\cite{SUZ1986}). Above 550~Hz no cloud of points is clearly identifiable in Fig.~\ref{fig:res_esprit_piano}(a) owing to a too-high Signal-to-Noise Ratio ($\approx35$~dB). Note that the modal overlap~$\mu$\footnote{The modal overlap is the ratio between the half-power modal bandwidth and the average modal spacing.} is around 30\% at 150~Hz and reaches 70\% at 550~Hz.
\hspace{-1cm}

\subsection{Acoustical excitation}
\begin{figure}[!ht]
\centering
  \includegraphics[width=0.35\linewidth]{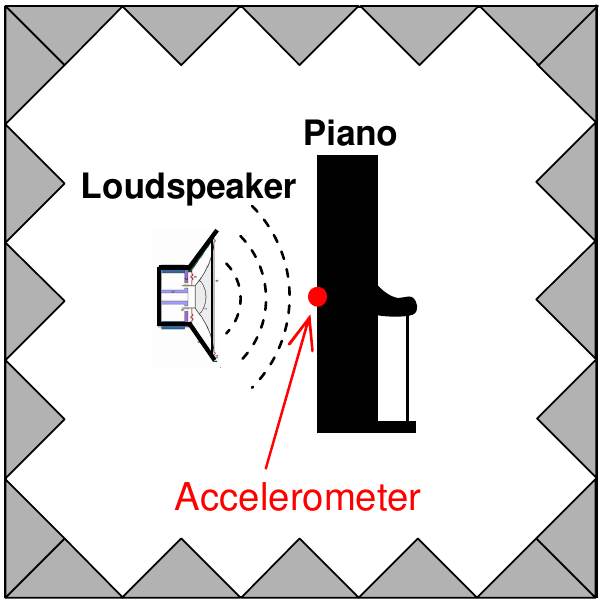}
  \caption{Acoustical excitation of the piano placed in a pseudo-anechoic room. The acceleration of the board is measured at four points.}
  \label{fig:piano_chambre}
\end{figure}

In order to improve the SNR and thus extend the estimation of the modal parameters towards higher frequencies (higher modal overlap, in fact), we have replaced the impulsive mechanical excitation by a continuous acoustical one (Fig.~\ref{fig:piano_chambre}).

\begin{figure}[!ht]
\centering
  \includegraphics[width=1\linewidth]{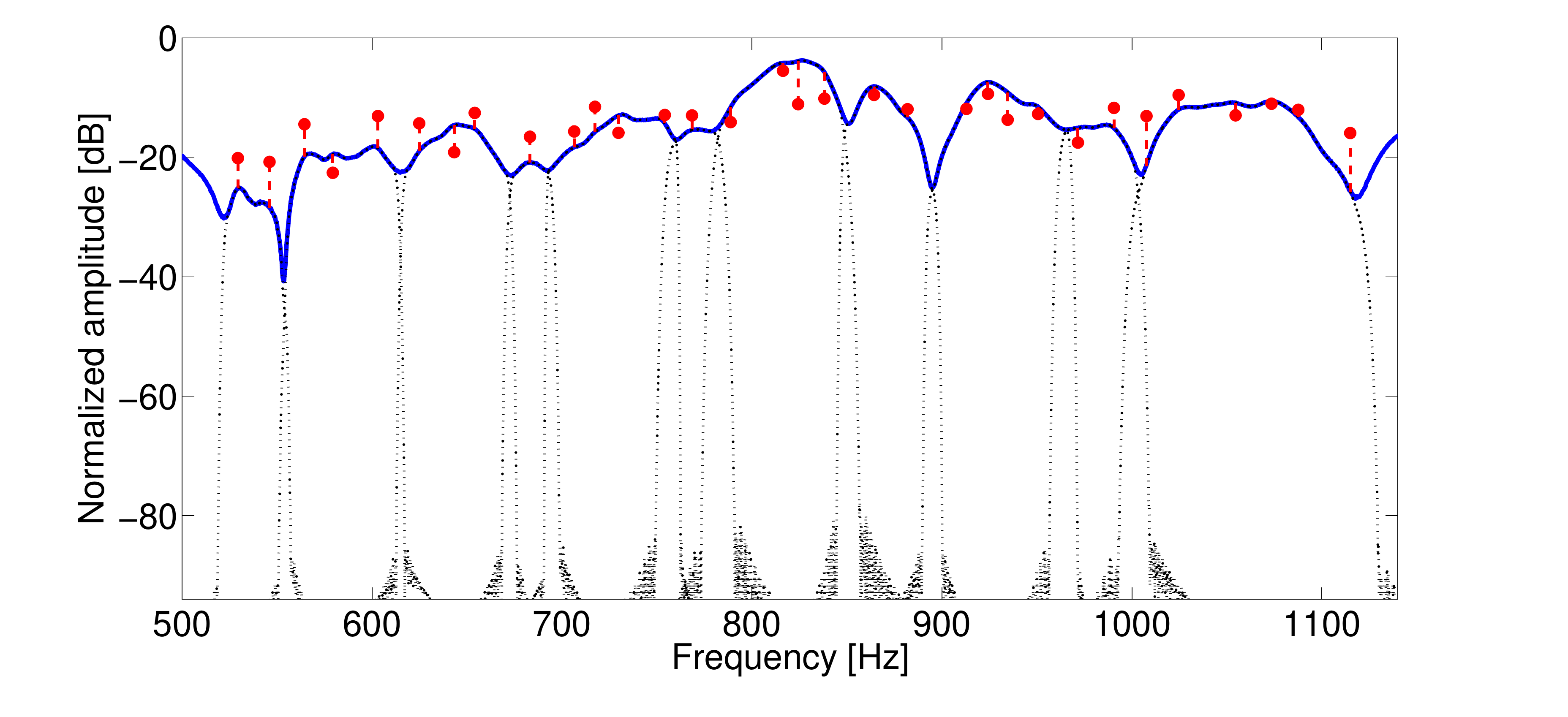}
  \caption{Typical bank-filtering analysis of a reconstructed impulse response between 550 et 1150~Hz (acoustical excitation). {\color[rgb]{0,0,1}---}\,: Fourier spectrum of the impulse response at point \textbf{A}$_\mathbf{2}$. {\color[rgb]{1,0,0}$\bullet\,$}: modes estimated by ESPRIT (modal amplitudes and frequencies). $\cdots\,$:~amplitude responses of the narrow pass-band filters.}
  \label{fig:spectre_500_1200_compoOK}
\end{figure}

The excitation of the loudspeaker is the same as in section~1 (a logarithmic swept-sine [50-4000]~Hz with a 40 kHz sampling frequency and a $T=26~\text{s}$ duration). The impulse response of the board is reconstructed by the deconvolution technique and analysed with a filter bank (a typical bank filtering analysis is displayed in Fig.~\ref{fig:spectre_500_1200_compoOK} between 550 and 1150~Hz). The cutoff frequencies of the finite-impulse-response (FIR) filters were chosen at local minima of the Fourier spectrum of the response. If necessary, when there is a doubt on the number of components in one frequency-band, two successive filters were occasionally chosen to overlap.

\begin{figure}[!ht]
\centering
  \includegraphics[width=1\linewidth]{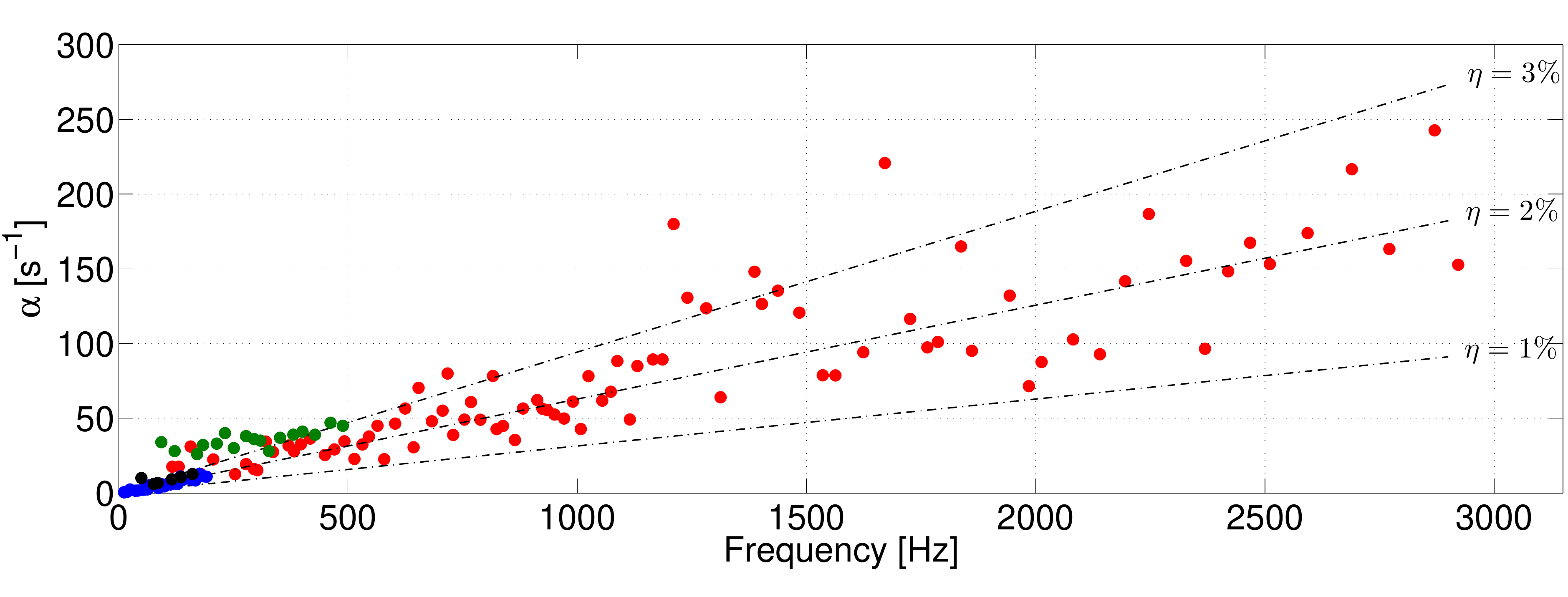}
  \caption{({\color[rgb]{1,0,0}{$\bullet$}}) : Modal damping factors estimated at point~\textbf{A}$_\mathbf{2}$ up to 3~kHz. Comparison with bibliographical results: {{$\bullet$}}~\cite{SUZ1986}; {\color[rgb]{0,0.5,0}{$\bullet$}}~\cite{DER1997}; {\color[rgb]{0,0,1}{$\bullet$}}~\cite{BER2003}.}
  \label{fig:res_esprit_amort_500_30008_compar}
\end{figure}

The modal damping factors as identified up to 3~kHz are reported in Fig.~\ref{fig:res_esprit_amort_500_30008_compar}, together with available bibliographical results (limited to $\approx500~Hz$). The advantage of the acoustical excitation technique in terms of frequency spanning is obvious.

Up to around 1200~Hz, loss factors range from 1\% to 3\% (mean of $\eta\approx2.3\%$ for the 55 estimations, lowest in frequency). Between 1200 and 1500~Hz, damping increases from a mean value of $\approx80~\text{s}^{-1}$ below 1200~Hz to $\approx130~\text{s}^{-1}$ in this domain. This increase can be attributed to the acoustical radiation of the structure since this frequency-band is the \textit{critical domain} of the soundboard where modes radiate most efficiently. It is interesting to note that these results (obtained on an upright piano) coincide with the ones obtained by Suzuki~\cite{SUZ1986} on a small grand of which he estimated the critical domain around 1400~Hz. Above 1.8~kHz the loss factors are of the order of the internal losses of spruce and influence of radiation is no more visible. A reason for this may be the alteration of the acoustical coincidence phenomenon at those high-frequencies, due to the localisation of the waves between the ribs (see last section). 

\subsection{Modal density}
\begin{figure}[ht!]
\begin{center}
\includegraphics[width=1\linewidth]{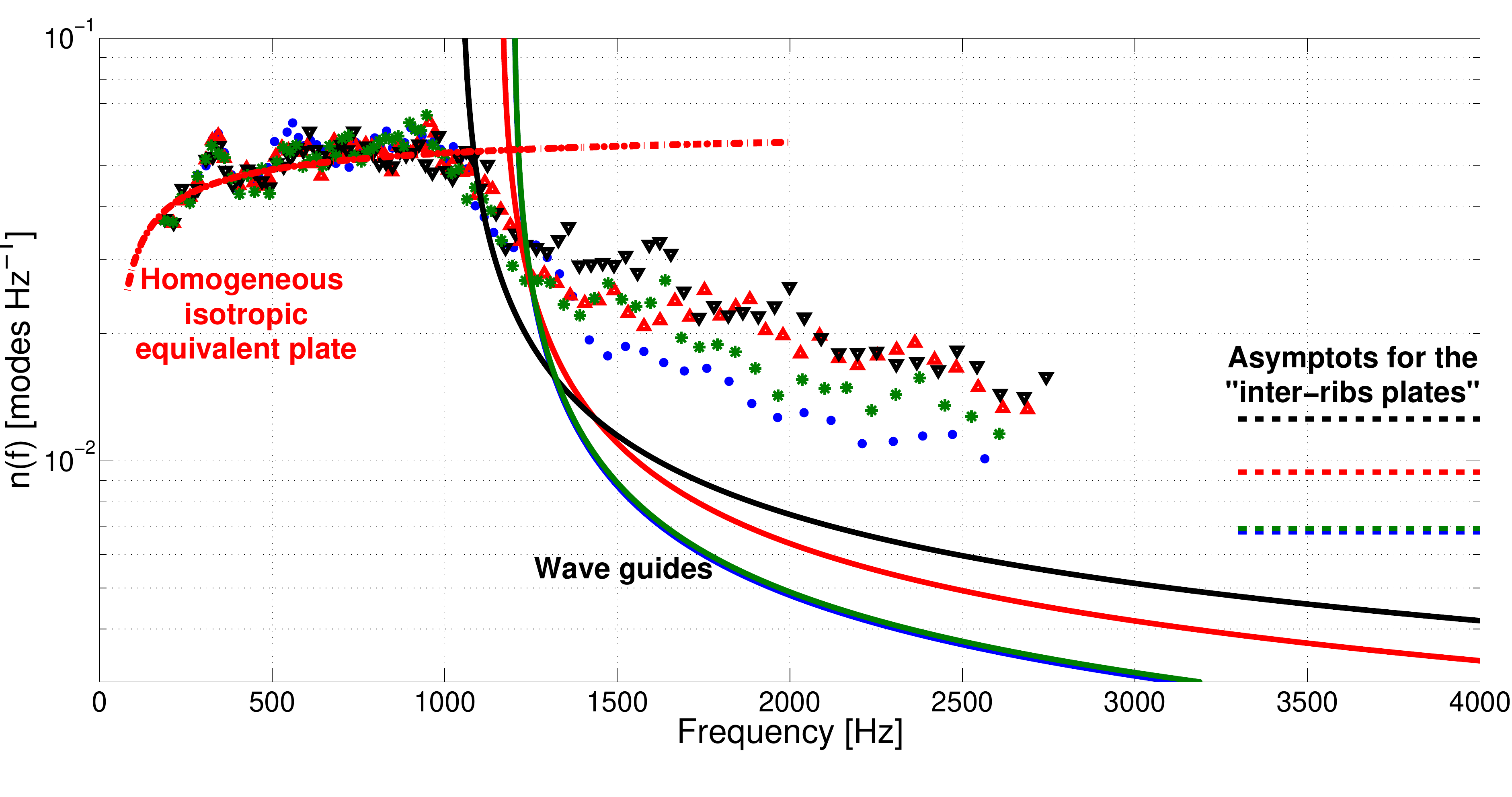}
\caption{Modal densities of the board measured at points \textbf{A}$_\mathbf{1}$~({\color[rgb]{0,0,1}\tiny{$\bullet$}}), \textbf{A}$_\mathbf{2}$~({\color[rgb]{1,0,0}\tiny{$\Delta$}}), \textbf{A}$_\mathbf{3}$~({\tiny{$\nabla$}}), \textbf{A}$_\mathbf{5}$~({\color[rgb]{0,0.5,0}\scriptsize{$\ast$}}) and calculated for the first waveguide mode (1,$n$) of the corresponding inter-rib space (---, with the corresponding color). Asymptotic values of the modal densities of the inter-rib plates (--~--, with the corresponding color). Theoretical modal density of the homogeneous isotropic equivalent clamped plate ({\color[rgb]{1,0,0}--$\,\cdot\,$--}), see next section.}
\label{fig:densitemodale_guidedonde}
\end{center}
\end{figure}

The modal density $n$ is a global descriptor of the vibratory behaviour of the structure in the mid- and high-frequency domain. It is given as a function of frequency in Fig.~\ref{fig:densitemodale_guidedonde} at four points of measurements (see Fig.~\ref{fig:table_exp_maillage} for the exact locations), as derived from an estimation of the modal spacing (moving average, six successive modes retained for each estimation). The frequency evolution of $n$ reveals two well-separated vibratory regimes of the structure.
\begin{enumerate}
\item{Below $1.1$~kHz, the 4 experimental curves are almost similar. $n(f)$ raises slowly and tends to a constant value of $\approx0.05~\text{modes}~\text{Hz}^{-1}$ independently of the zones of the board where the measure is done. \textbf{The ribbed board behaves as a homogeneous isotropic plate} (see next section). The slow rise confirms moreover that the boundary conditions are \emph{constraint}.}
\item{For frequencies above $1.1$~kHz, $n(f)$ decreases significantly. Ribs confine the wave propagation: \textbf{the soundboard behaves as a set of waveguides} (see last section).}
\end{enumerate}

\section{Modelling the low-frequency behaviour: a thin homogeneous plate in an isotropic equivalent material}
\subsection{A simple finite-element model}
The modal density of the soundboard below 1.1~kHz suggests that a homogeneous isotropic plate represents a good model in this frequency domain. In order to confirm this numerically, we realised a two-dimensional finite-element model (FEM) of the soundboard by means of the free software CAST3M (triangular elements of 2~cm). The ribbed zone was replaced by an isotropic plate with a dynamical rigidity equal to the one of the orthotropic spruce plate in the direction of the grain:
\begin{equation}
D^{H}/(\rho^H\,h^H)=D_x^{p}/(\rho^p\,h^p)=156~\text{m}^4~s^{-2}
\end{equation}
where the constants of rigidity are generically $D=Eh^3/(12(1-\nu^2))$, with the Young's modulus $E$ and the Poisson's ratio $\nu$. The plate density and thickness are $\rho$ and $h$ respectively. The superscript $^\text{p}$ refers to the unribbed orthotropic spruce plate and $^\text{H}$ refers to the homogeneous equivalent isotropic plate. The $x$-axis corresponds to the direction of the grain of the wood, orthogonal to the ribs (see Fig.~\ref{fig:table_exp_maillage}). In the numerical model, the two \emph{dead zones} in the upper right and bottom left corners of the soundboard are modelled as orthotropic spruce plates with a constant thickness of $8$~mm. The two bars delimiting these two zones are made of fir. The mechanical characteristics used in the numerical model for spruce and fir are summarised in Table~\ref{tab:caracmeca_num}, as derived from measurements made by Berthaut~\cite{BER2004} on spruce and fir species selected for piano soundboards.
\begin{table}[!ht]
\begin{center}
\begin{tabular}{|c||c|c|c|c|c|c|c|c|c|c|c|}
\hline
\  & $E_\text{L}$& $E_\text{R}$ & $G_\text{LR}$& $\nu_\text{LR}$ & $\rho$~[kg~m$^{-3}$]\\
\hline

\hline
Spruce & 11.5 & 0.47 & 0.5 & 0.005 & 392 \\
\hline
Fir & 8.86 & 0.54 & 1.6 & 0.005 & 691\\
\hline
\end{tabular}
\end{center} 
\caption{Mechanical characteristics of spruce and fir species selected for piano soundboard, after~\cite{BER2004}. The longitudinal and radial Young's moduli ($E_\text{L}$ and $E_\text{R}$) and shear modulus ($G_\text{LR}$) are given in GPa.}
\label{tab:caracmeca_num}
\end{table}

\subsection{Results}
\begin{figure}[!ht]
\centering
\includegraphics[width=1\linewidth]{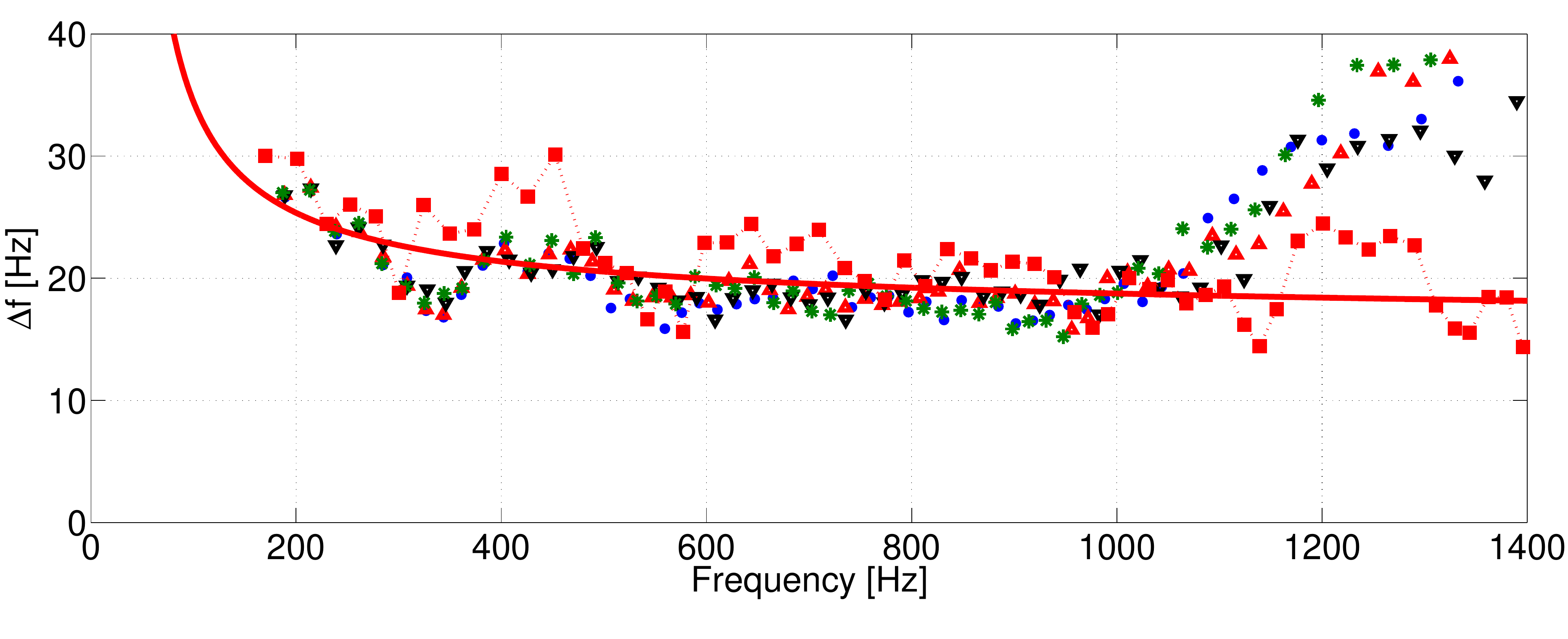}
  \caption{Average modal spacing of the piano soundboard: measured at points\textbf{A}$_\mathbf{1}$~({\color[rgb]{0,0,1}\tiny{$\bullet$}}), \textbf{A}$_\mathbf{2}$~({\color[rgb]{1,0,0}\tiny{$\Delta$}}), \textbf{A}$_\mathbf{3}$~({\tiny{$\nabla$}}), \textbf{A}$_\mathbf{5}$~({\color[rgb]{0,0.5,0}\scriptsize{$\ast$}}), numerically calculated with a finite-element model of a homogeneous isotropic clamped plate ({\color[rgb]{1,0,0}\tiny{$\blacksquare$}}) and theoretical for the same plate ({\color[rgb]{1,0,0}---}).}
  \label{fig:compar_plaque_isotrope}
\end{figure}

\begin{figure}[!h]
\begin{center}
\subfigure[(1,1)-mode]
     {\includegraphics[width=0.32\linewidth]{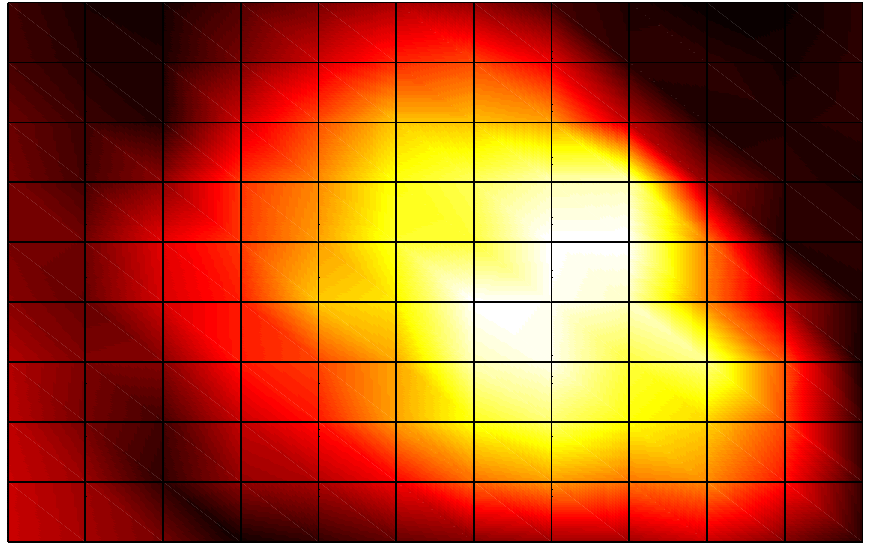}}
\subfigure[(2,1)-mode]
     {\includegraphics[width=0.32\linewidth]{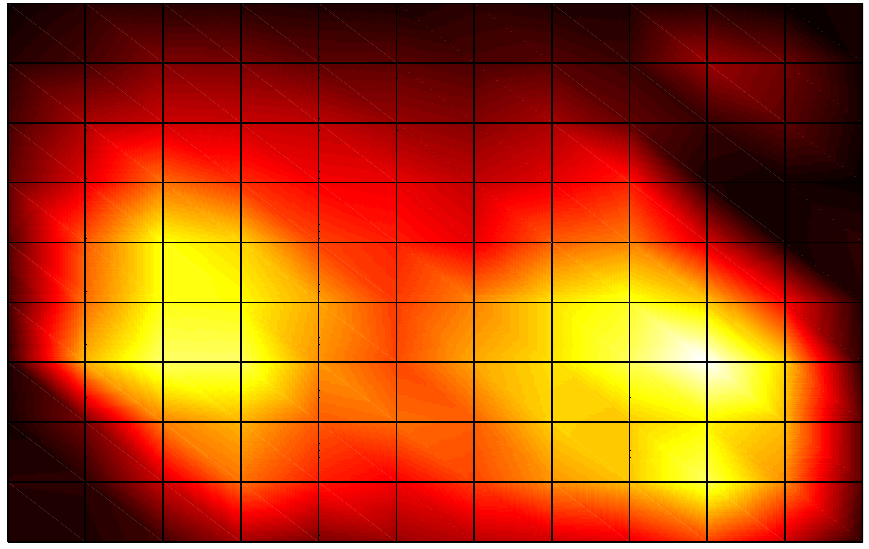}}
\subfigure[(3,1)-mode]
     {\includegraphics[width=0.32\linewidth]{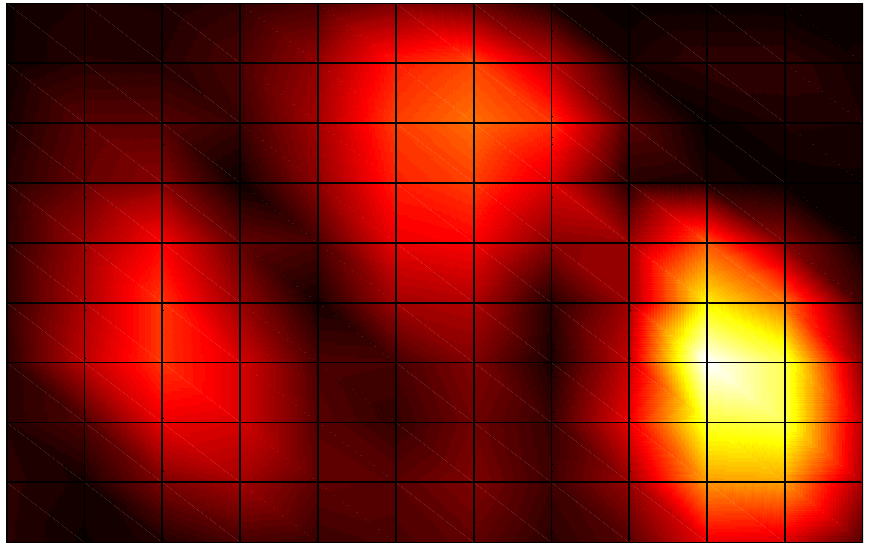}}\\
\subfigure
	 {\includegraphics[angle=90,width=0.32\linewidth]{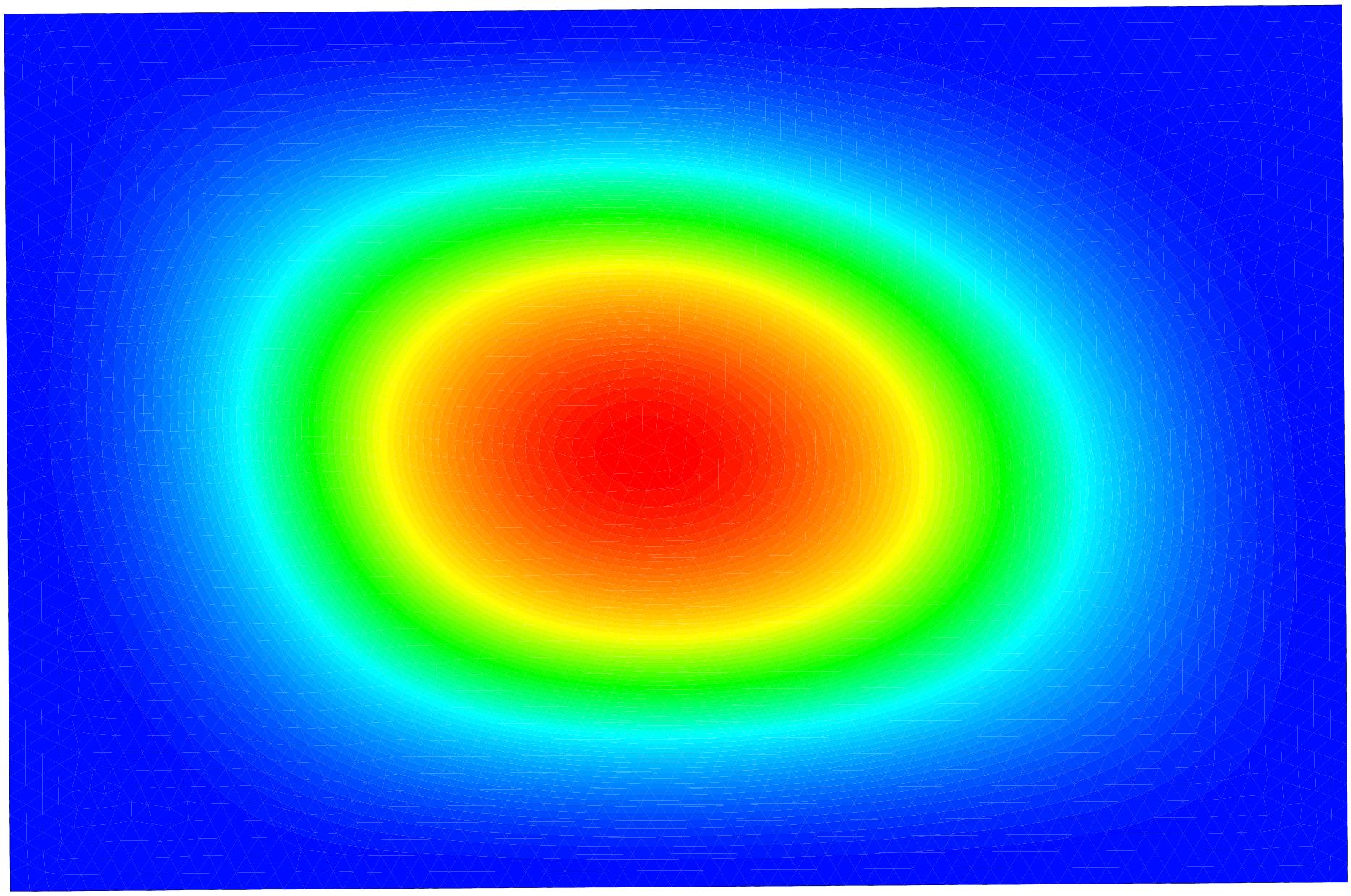}}
\subfigure
     {\includegraphics[angle=90,width=0.32\linewidth]{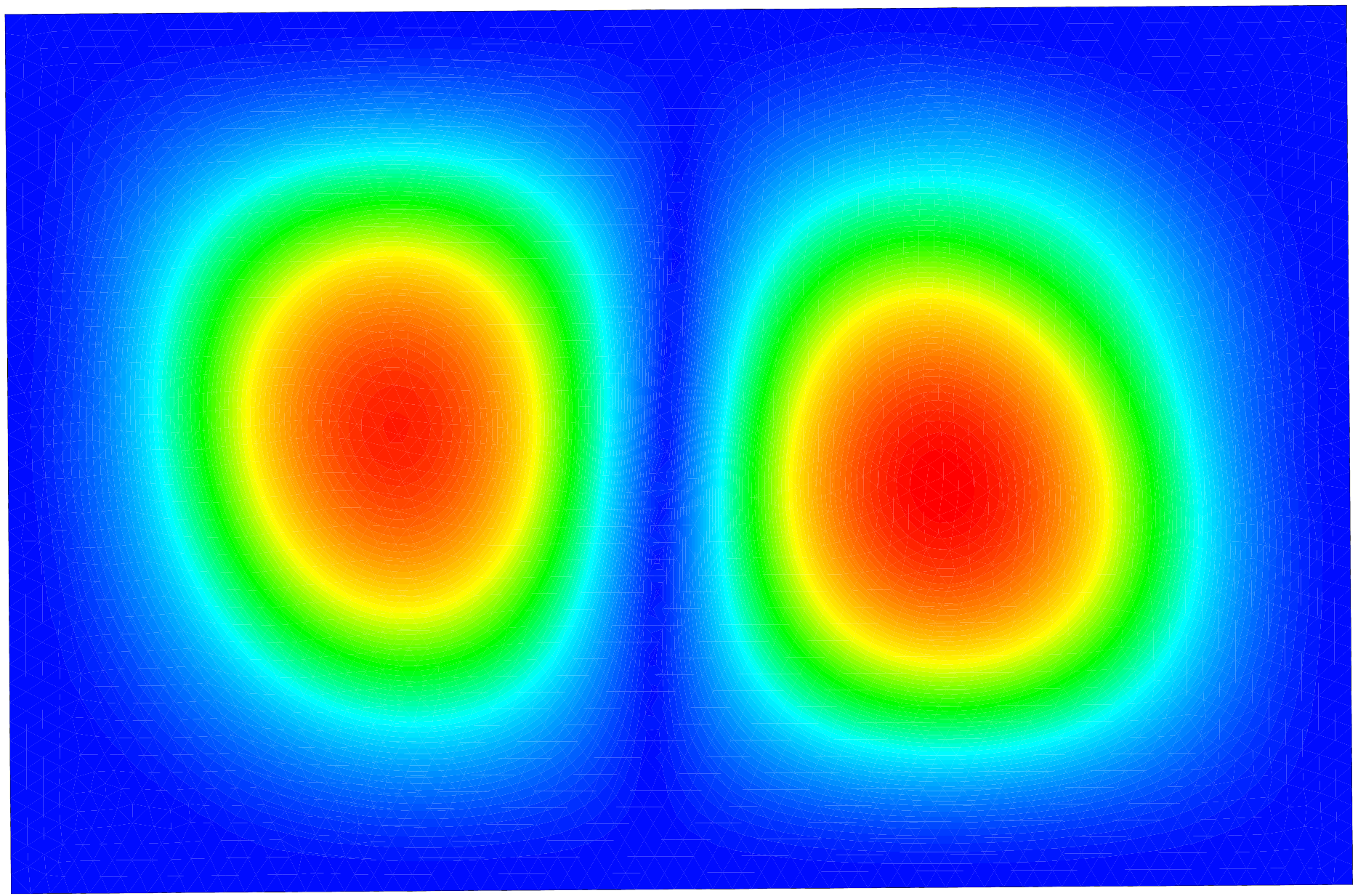}}
\subfigure
     {\includegraphics[angle=90,width=0.32\linewidth]{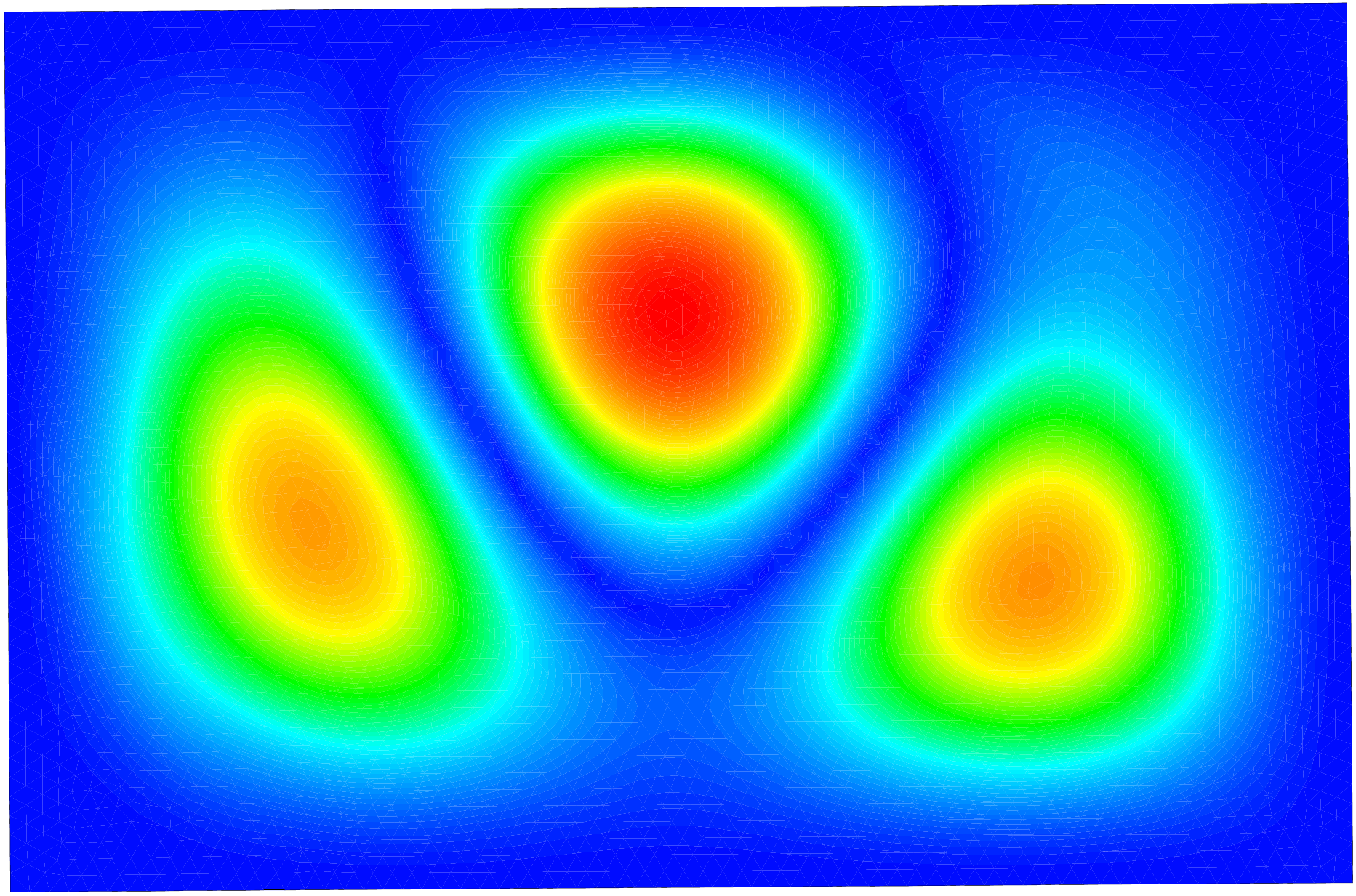}}\\
     \subfigure
   {\includegraphics[angle=90,width=0.32\linewidth]{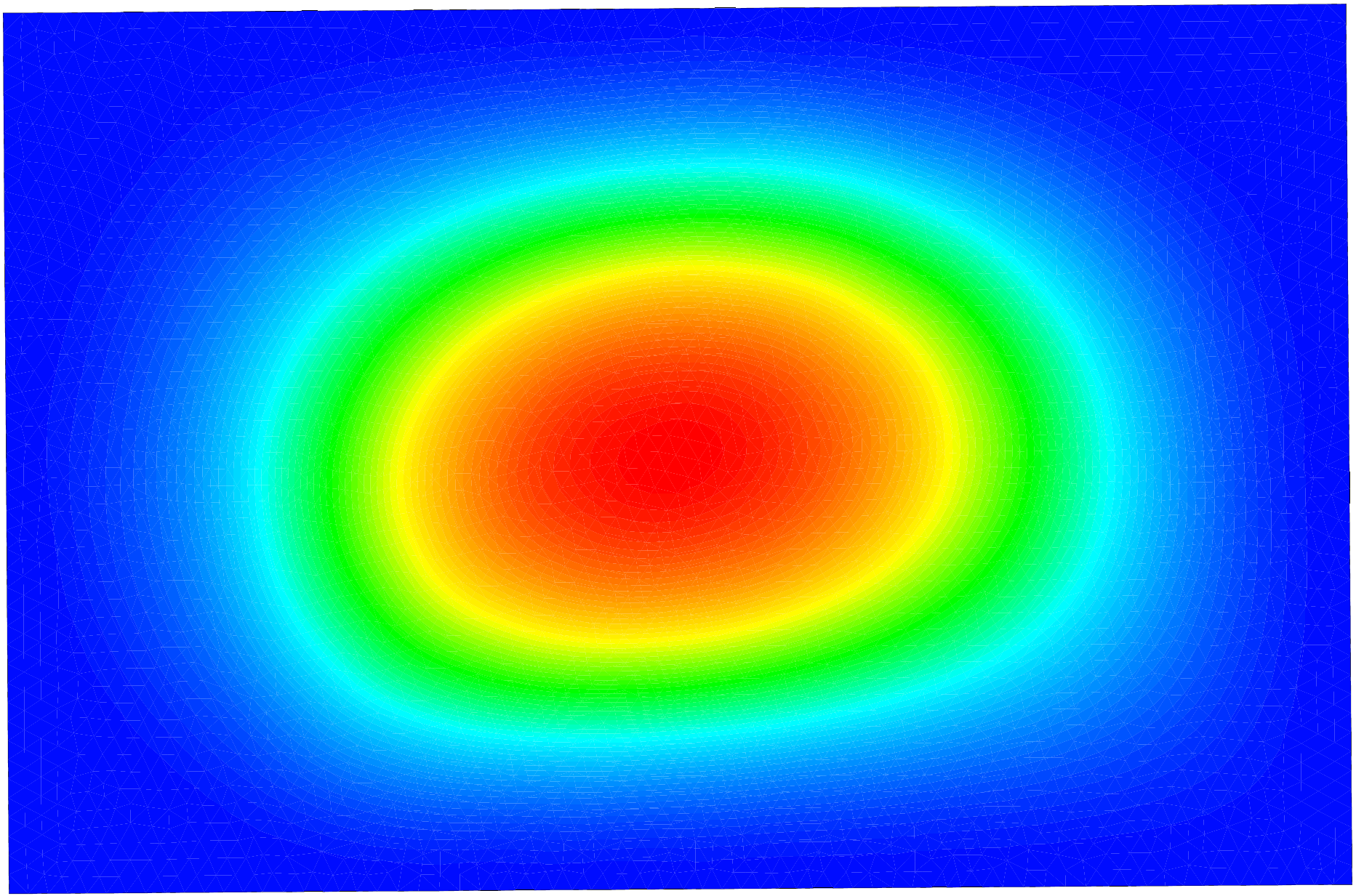}}
\subfigure
    {\includegraphics[angle=90,width=0.32\linewidth]{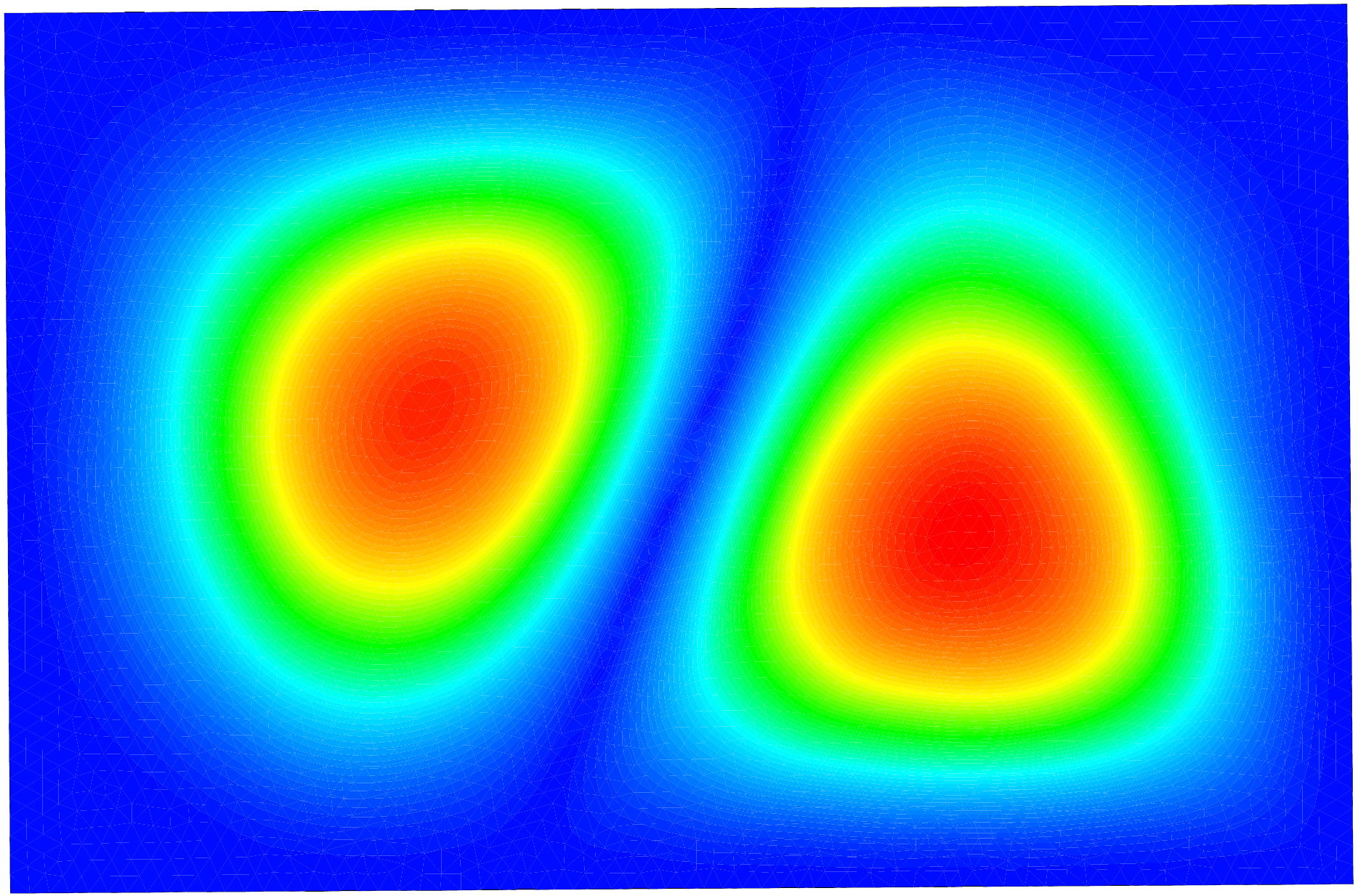}}
\subfigure
     {\includegraphics[angle=90,width=0.32\linewidth]{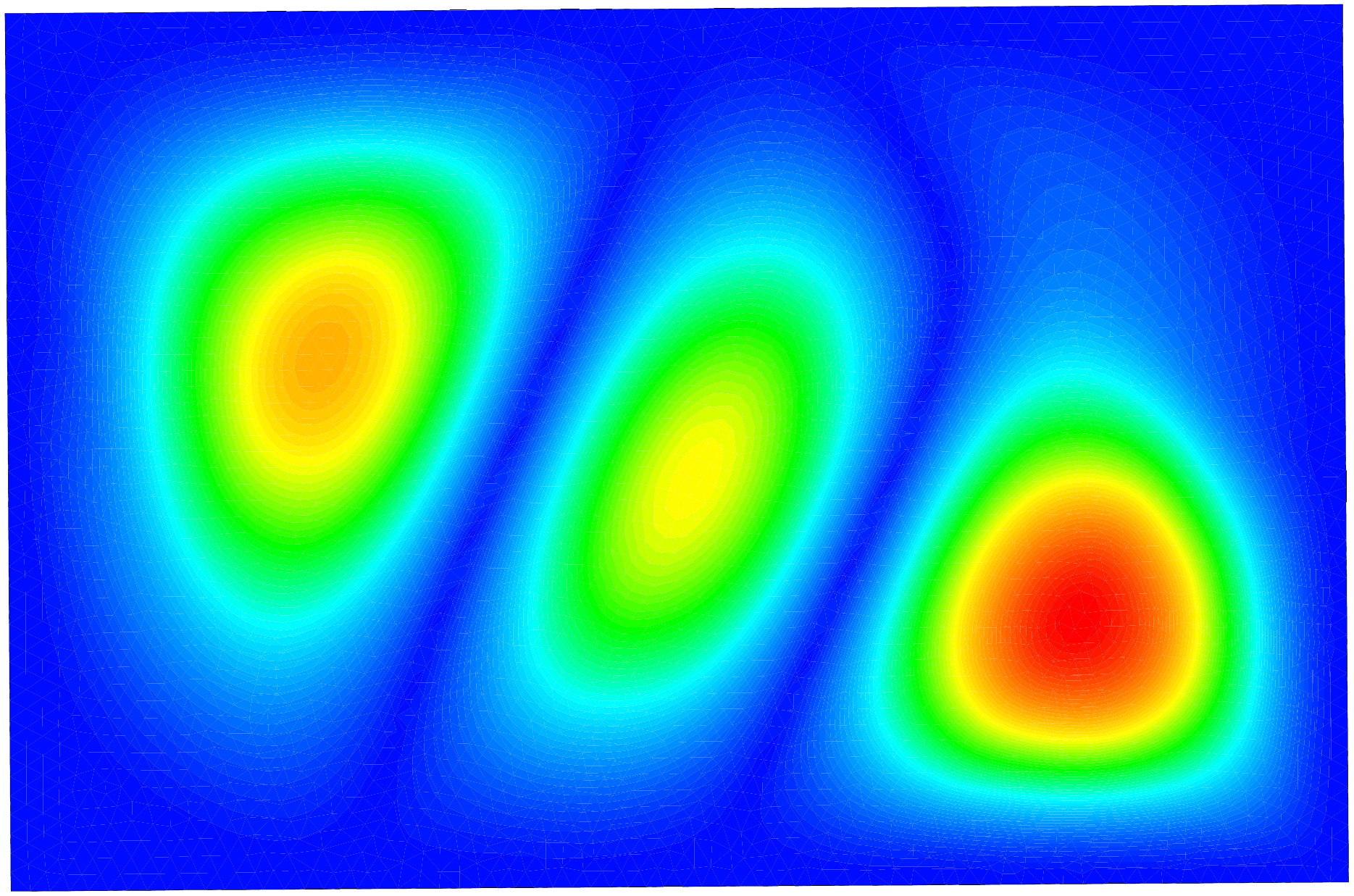}}
\caption{First three modal shapes of the upright piano soundboard: measured (first line), numerical (FEM) after replacement of the ribbed zone by a homogeneous isotropic equivalent plate (middle line), numerical (FEM) for the ribbed board without bridges (bottom line).}
 \label{fig:modes_castemiso}
\end{center}
\end{figure}

The numerical solution of the eigenvalue problem is obtained in the conservative case and for \emph{clamped} boundary conditions. The numerical average modal spacings is presented in Fig.~\ref{fig:compar_plaque_isotrope} together with the experimental estimations. The theoretical average modal spacing of an isotropic clamped plate is also included. Numerical results match well the experimental estimations in the frequency domain of interest. Moreover the first numerical modal shapes for the isotropic plate -- presented in Fig.~\ref{fig:modes_castemiso} -- are remarkably similar to those obtained experimentally, whereas the one obtains with a numerical model of the ribbed board \emph{without bridges} are \emph{less similar}. This confirms that on a typical piano soundboard, the ribs and the bridges compensate globally the anisotropy of spruce.

\begin{figure}[!ht]
\centering
\includegraphics[width=0.8\linewidth]{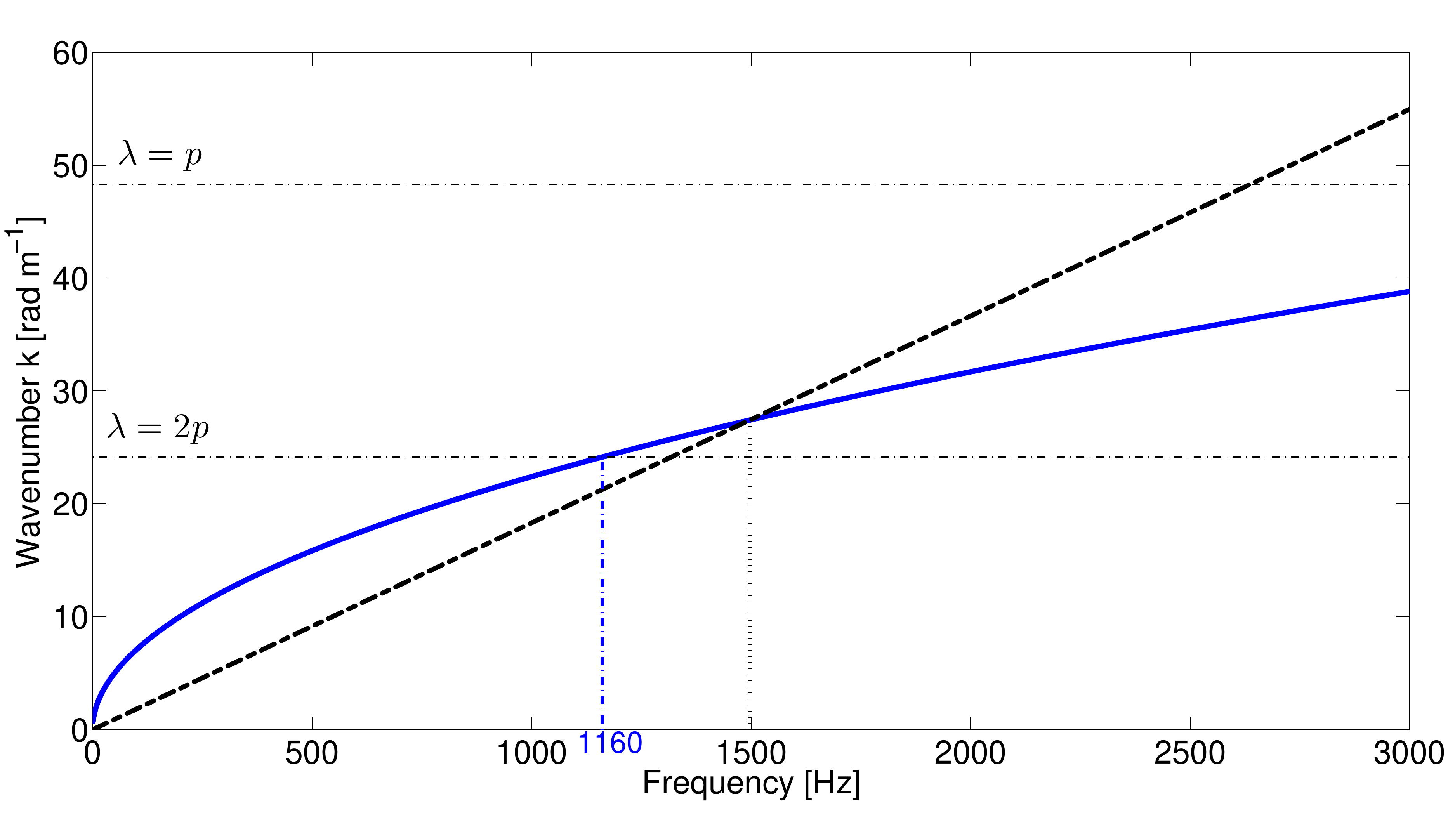}
  \caption{Dispersion curves in the isotropic plate ({\color[rgb]{0,0,1}---}), and in air (--~--). The critical frequency of radiation of the plate is $\approx1500$~Hz. At $\approx1160$~Hz, the half-wavelength $\lambda/2$ is equal to the average distance $p$ between two consecutive ribs.}
  \label{fig:dispers}
\end{figure}

To conclude this section, we reproduce on figure~\ref{fig:dispers} the relation of dispersion of flexural waves in the isotropic plate. The coincidence frequency (for which the dispersion curve in the plate intersects the one in air) is $\approx1500$~Hz which is consistent with previous experimental conclusions. However, the half-wavelength $\lambda/2$ in the plate become equal or less to the average distance $p$ between two consecutive ribs ($\approx13$~cm) at $\approx1160$~Hz so that beyond this frequency, the structure cannot be modelled as a homogeneous plate any more (see next section).

\section{Modelling the mid- and high-frequency behaviour: set of waveguides and modified coincidence phenomenon}

\subsection{Numerical observations}
\begin{figure}[!h]
\begin{center}
\subfigure[32$^\text{nd}$ numerical modal shape, $f_\text{n}=776~Hz$]
{\includegraphics[width=0.45\linewidth]{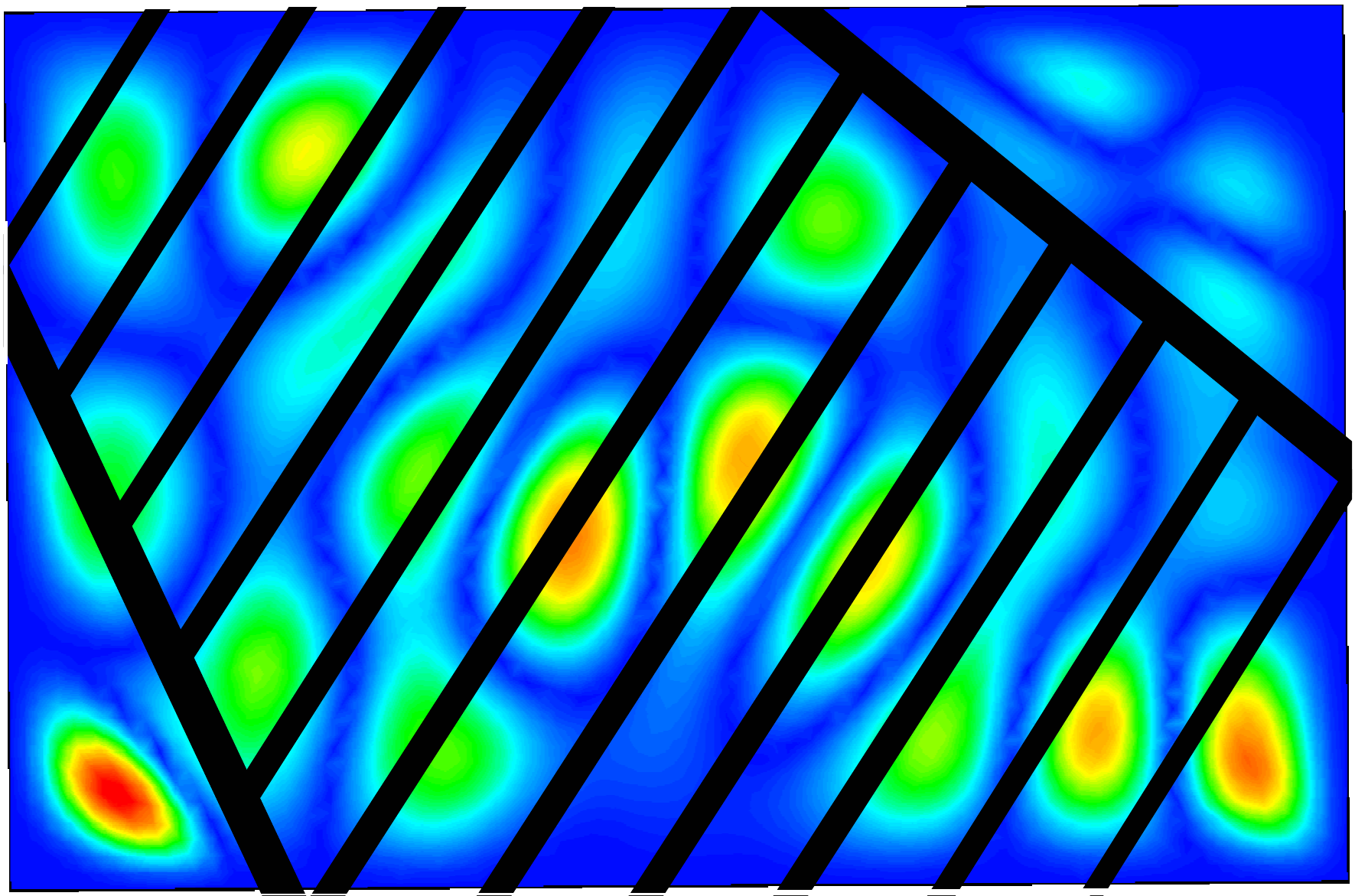}}
   \hspace{0.04\linewidth}
\subfigure[48$^\text{th}$ numerical modal shape, $f_\text{n}=1089~Hz$]
{\includegraphics[width=0.45\linewidth]{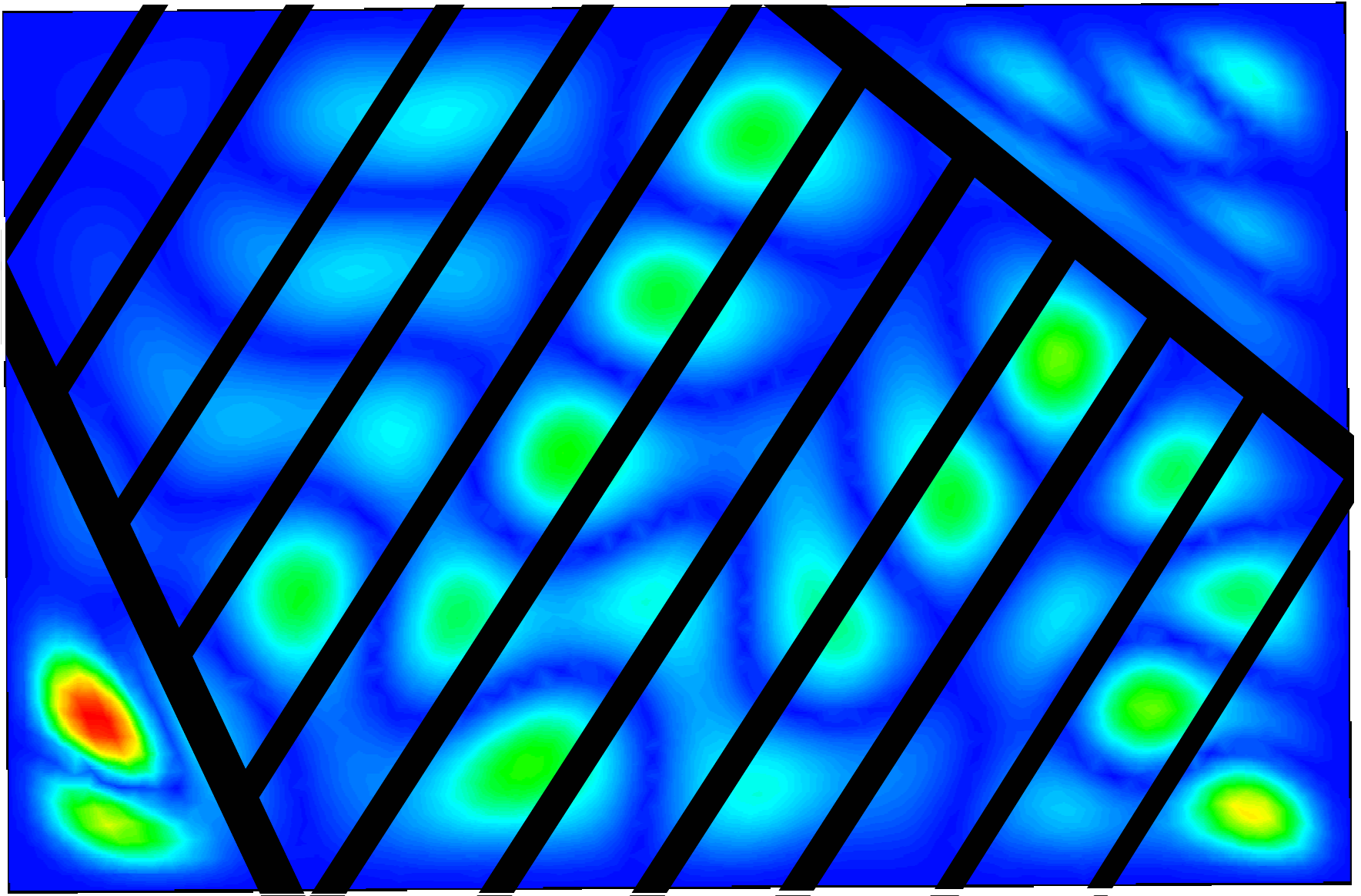}}\\
\subfigure[74$^\text{th}$ numerical modal shape, $f_\text{n}=1593~Hz$]
{\includegraphics[width=0.65\linewidth]{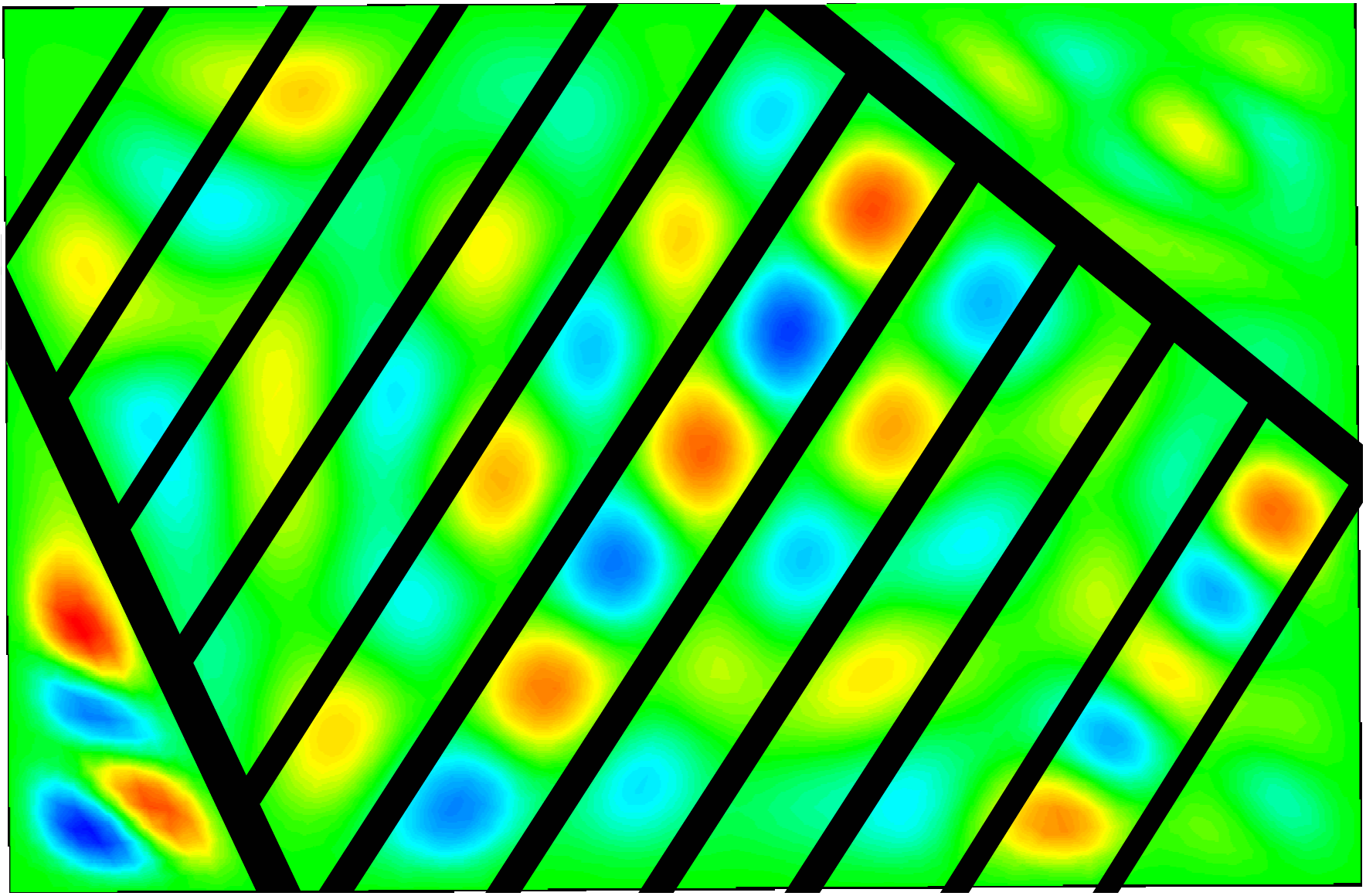}}
\caption[]{Modal shapes obtained by a FEM of the ribbed board. The positions of the ribs and of the two bars are indicated in black. At 1593~Hz, the antinodes of vibration are localised between the ribs (nodes of vibration). This behaviour is already visible for the modal shapes at 1089~Hz, but not at 776~Hz. In (c), the sign of the phase is represented: red and blue zones denote regions vibrating with opposite phases.}
 \label{fig:castem_HF}
\end{center}
\end{figure}

For frequencies above 1.1 kHz, $n(f)$ falls significantly. At 1.1 kHz, the transverse waves in a soundboard without ribs would have a half-wavelength equal to the average distance $p$ between two consecutive ribs. Berthaut~\emph{et al.}~\cite{BER2004} made the observation that ribs confine the wave propagation. This is confirmed here by numerical simulations. Modal shapes obtained through the finite-element model of the ribbed board~(Fig.~\ref{fig:castem_HF}) exhibit a localisation of the waves for frequencies above 1.1~kHz. The soundboard behaves as a set of waveguides.

\subsection{The waveguide model}
\begin{figure}[ht!]
\begin{center}
\includegraphics[width=0.6\linewidth]{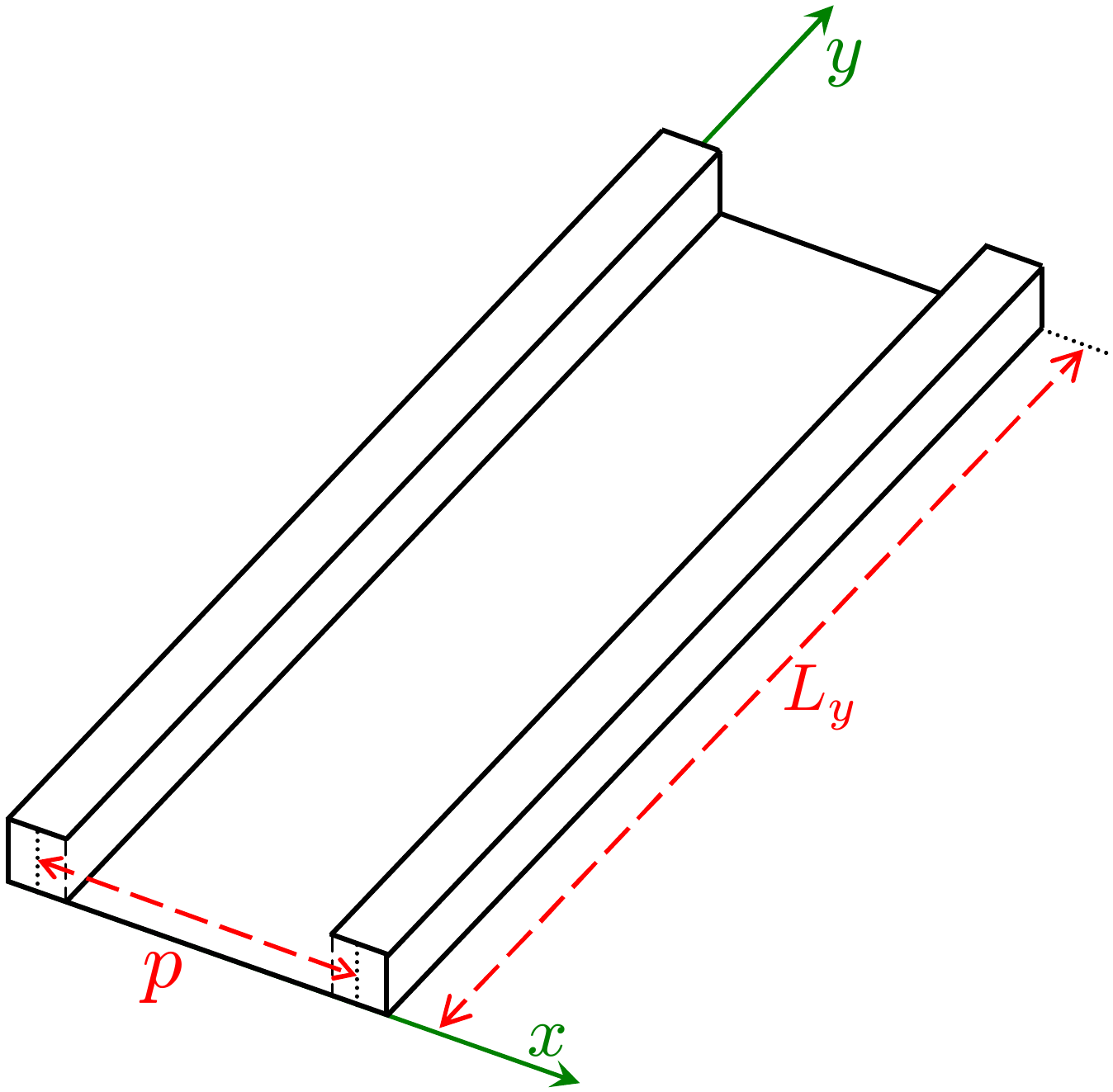}
\caption{The board between two consecutive ribs can be considered in the high-frequency range as a \emph{finite} waveguide.}
\label{fig:guidedonde}
\end{center}
\end{figure}

We adopt the simplest possible waveguide model (Fig.~\ref{fig:guidedonde}) in order to derive a dispersion law of this non-homogeneous plate and to calculate its modal density.

The hypotheses are~:
\begin{itemize}
\item{The inter-rib region behaves like an orthotropic plate with clamped boundaries.}
\item{Wavenumbers $k_{x_m}$ in the direction $x$ normal to the ribs\footnote{The direction $x$ is parallel to the grain of the spruce board and, by construction, perpendicular to the ribs (see figure~\ref{fig:table_exp_maillage}).} are imposed by the inter-rib distance $p$: $k_{x_m}=m\pi/p$ with $m\in\mathbb{N}^*$.}
\item{Wavenumbers $k_{y_n}$ in the direction $y$ parallel to the ribs are imposed by the boundary conditions at rim.}
\end{itemize}

According to the first hypothesis, the dispersion law is:
\begin{equation}\label{eq:dispersguide}
k_y^4+k_y^2\,\cfrac{D_2+D_4}{D_3}\,k_{x_m}^2+\cfrac{D_1}{D_3}\,k_{x_m}^4-\cfrac{\rho\,h\,\omega^2}{D_3}=0
\end{equation}
where the $D_i$ are the constants of rigidity of spruce, considered as an orthotropic material (of main axes $x$ and $y$):\\
$D_1=E_x h^3/(12(1-\nu_{xy}\nu_{yx}))$, $D_2=\nu_{yx}E_x h^3/(6(1-\nu_{xy}\nu_{yx}))$,\\ $D_3=E_yh^3/(12(1-\nu_{xy}\nu_{yx}))$ and $D_4=G_{xy}h^3/3$. 

With $A=\cfrac{D_2+D_4}{D_3}\,k_{x_m}^2$, $B=\cfrac{D_1}{D_3}\,k_{x_m}^4$ and $C=\cfrac{\rho\,h}{D_3}$, the equation \ref{eq:dispersguide} becomes a second order equation in $k_y^2$: $k_y^4+A\,k_y^2+B-C\,\omega^2=0$.  Finally the dispersion law in each waveguide is:
\begin{equation}
k_y=\pm\left(\cfrac{\sqrt{A^2+4C\omega^2-4B}-A}{2}\right)^{1/2}
\end{equation}
The pulsation $\omega_{\text{c},p,m}=\sqrt{B/C}=\left(\cfrac{m\pi}{p}\right)^2\sqrt{\cfrac{D_1}{\rho\,h}}$ is a low cutoff pulsation for the transverse modes~$m$ in ribs separated by $p$. Below this pulsation, there is no real solution and waves are evanescent in the $y$ direction.

For a given $k_{x_m}$, the modal density in the wave guide can be calculated as follows. The wavenumbers $k_{y_n}$ are approximated by $n\pi/L_y$~with $n\in\mathbb{N}^*$. The number of modes of pulsation less than $\omega$ is $N(\omega)=k_y(\omega)L_y/\pi\,$ and the modal density:
\begin{equation}
n(\omega)=\cfrac{\text{d}N}{\text{d}k_y}\,\cfrac{\text{d}k_y}{\text{d}\omega}
=\cfrac{L_y}{\pi}\,\cfrac{\sqrt{2}\,C\omega}{\sqrt{A^2+4C\omega^2-4B}\:\left(\sqrt{A^2+4C\omega^2-4B}-A\right)^{1/2}}
\end{equation}

In the high-frequency limit, the waveguide has the same modal density as that of a beam of length $L_y$, with a $\omega^{-1/2}$ dependency:
\begin{equation}
n(\omega)\:\underset{\omega\rightarrow +\infty}{\rightarrow}\:\cfrac{L_y}{\pi}\,\cfrac{C^{1/4}}{2\,\sqrt{\omega}}=\cfrac{L_y}{2\pi\sqrt{\omega}}\,\left(\cfrac{\rho\,h}{D_3}\right)^{1/4}
\end{equation}

The modal density of waveguides with different values of $p$ and $L_y$ are reported in Fig.~\ref{fig:densitemodale_guidedonde}.

\subsection{The modified acoustical coincidence}
\begin{figure}[ht!]
\begin{center}
\includegraphics[width=1\linewidth]{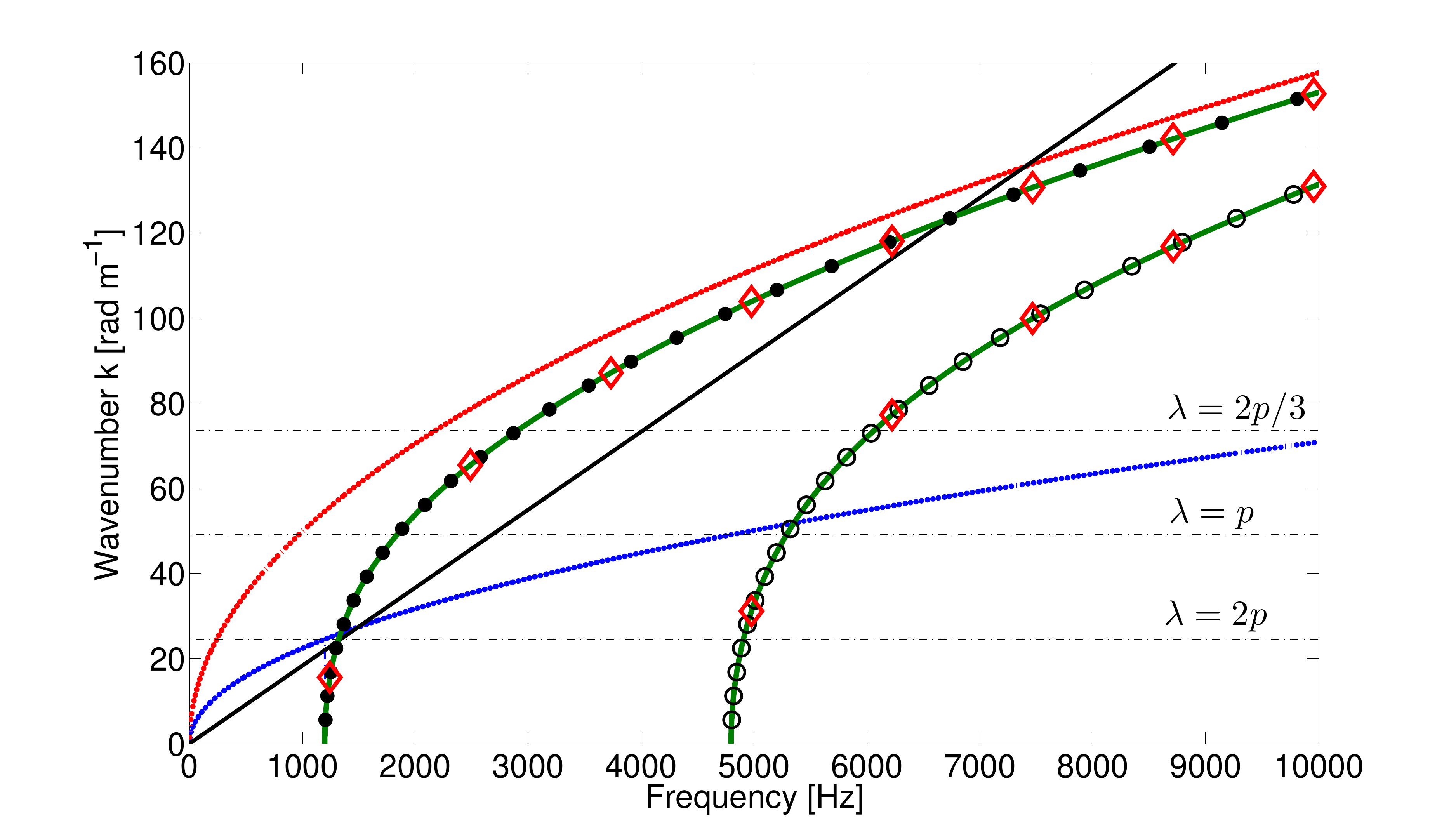}
\caption{Relations of dispersion for flexural waves in the orthotropic plate ({\color[rgb]{0,0,1}{$\cdots$}} along~$x$ and {\color[rgb]{1,0,0}{$\cdots$}} along~$y$), in the air (---) and for the two first waveguide modes ({\color[rgb]{0,0.5,0}{---}} and {\color[rgb]{0,0.5,0}{--~--}}) of the waveguide between the second and third ribs. We add the discretes values corresponding to the waveguide modes $(1,n)$ in $\bullet$ and $(2,n)$ in ($\circ$), together with the supposed perfectly harmonic partials of the~$\mathbf{\textbf{D}\sharp_6}$ strings in {\color[rgb]{1,0,0}$\diamond$}.}
\label{fig:dispers_guide}
\end{center}
\end{figure}

The acoustical coincidence phenomenon is deeply modified in comparison with the one occurring in a thin plate (see figure~\ref{fig:dispers_guide}). The dispersion curve of a waveguide can present one, two, or no coincidence frequencies depending on the value of $p$. This creates a nonuniformity in the radiation of the soundboard in the treble range of the instrument compared to the lower range and this may explain the difference in timbre. For example, for the key $\mathbf{\textbf{D}\sharp_6}$ having a fundamental frequency around 1245~Hz, the damping factor due to the acoustical radiation of the fundamental may be higher (supersonic waves) than the damping factors of the next two partials (subsonic waves). 

\section{Conclusion}
We have investigated the vibrations of the soundboard of an upright piano in playing condition. At normal levels of vibration, the linear part of the mechanical response to an acoustical response is $\approx$50--60 dB above the nonlinear component. 
Given the essentially linear response, a modal identification was
performed in the mid-frequency domain [300-2500]~Hz by means of a novel high-resolution modal
analysis technique: the modal density and the loss factor could be measured up to 2.5--3~kHz. The frequency evolution of the modal density of the piano soundboard reveals two well-separated vibratory regimes of the structure. 
Below 1.1 kHz, the modal density is very close to that of a homogeneous isotropic plate with clamped boundary conditions. Higher in frequency, the soundboard behaves as a set of waveguides defined by the ribs. A numerical (FEM) determination of the modal shapes confirms that the waves are localised between the ribs. The measured modal density is consistent with an estimation based on the dispersion law of waves in each waveguide. The acoustical coincidence phenomenon is deeply modified in comparison with that occurring in thin plates. The dispersion curve of a waveguide can present one, two, or no coincidence frequencies. This creates a nonuniformity in the radiation of the soundboard in the treble range of the instrument compared to the lower range and this may explain the difference in timbre.

Finally, we would like to notify the reader that the measurements presented in this paper have been used to give a synthetic description of the piano soundboard mechanical mobility (admittance): see the ISMA companion-paper 
\cite{EGE2010_2}).

\bibliography{ica_ege_boutillon_soundboard}

\begin{thebibliography}{12}
\providecommand{\natexlab}[1]{#1}
\providecommand{\url}[1]{\texttt{#1}}
\expandafter\ifx\csname urlstyle\endcsname\relax
  \providecommand{\doi}[1]{doi: #1}\else
  \providecommand{\doi}{doi: \begingroup \urlstyle{rm}\Url}\fi

\bibitem[Askenfelt and Jansson(1992)]{ASK1992}
A.~Askenfelt and E.~V. Jansson.
\newblock On vibration sensation and finger touch in stringed instrument
  playing.
\newblock \emph{Music Perception}, 9\penalty0 (3):\penalty0 311--349, 1992.

\bibitem[Badeau et~al.(2006)Badeau, David, and Richard]{BAD2006}
R.~Badeau, B.~David, and G.~Richard.
\newblock A new perturbation analysis for signal enumeration in rotational
  invariance techniques.
\newblock \emph{IEEE Transactions on Signal Processing}, 54\penalty0
  (2):\penalty0 450--458, 2006.

\bibitem[Berthaut(2004)]{BER2004}
J.~Berthaut.
\newblock \emph{Contribution à l'identification large bande des structures
  anisotropes. Application aux tables d'harmonie des pianos}.
\newblock {PhD Thesis}, École Centrale de Lyon, 2004.

\bibitem[Berthaut et~al.(2003)Berthaut, Ichchou, and Jezequel]{BER2003}
J.~Berthaut, M.~N. Ichchou, and L.~Jezequel.
\newblock Piano soundboard: structural behavior, numerical and experimental
  study in the modal range.
\newblock \emph{Applied Acoustics}, 64\penalty0 (11):\penalty0 1113--1136,
  2003.

\bibitem[Dérogis(1997)]{DER1997}
P.~Dérogis.
\newblock \emph{Analyse des vibrations et du rayonnement de la table d'harmonie
  d'un piano droit et conception d'un système de reproduction du champ
  acoustique}.
\newblock {PhD Thesis}, Université du Maine, Le Mans, 1997.

\bibitem[Ege and Boutillon(2010)]{EGE2010_2}
K.~Ege and X.~Boutillon.
\newblock Synthetic description of the piano soundboard mechanical mobility.
\newblock In \emph{20th International Symposium on Music Acoustics}, Sydney and
  Katoomba, Australia, 2010.

\bibitem[Ege et~al.(2009)Ege, Boutillon, and David]{EGE2009}
K.~Ege, X.~Boutillon, and B.~David.
\newblock High-resolution modal analysis.
\newblock \emph{Journal of Sound and Vibration}, 325\penalty0 (4-5):\penalty0
  852--869, 2009.

\bibitem[Farina(2000)]{FAR2000}
A.~Farina.
\newblock Simultaneous measurement of impulse response and distortion with a
  swept-sine technique.
\newblock In \emph{108th Convention of the Audio Engineering Society}, Paris,
  France, 2000.

\bibitem[Roy and Kailath(1989)]{ROY1989}
R.~Roy and T.~Kailath.
\newblock {ESPRIT} - estimation of signal parameters via rotational invariance
  techniques.
\newblock \emph{IEEE Transactions on Acoustics Speech and Signal Processing},
  37\penalty0 (7):\penalty0 984--995, 1989.

\bibitem[Rébillat et~al.(2010)Rébillat, Hennequin, Corteel, and Katz]{REB2010}
M.~Rébillat, R.~Hennequin, E.~Corteel, and B.~F.~G. Katz.
\newblock Identification of cascade of hammerstein models for the description
  of non-linearities in vibrating devices.
\newblock \emph{Submitted to the Journal of Sound and Vibration}, 2010.

\bibitem[Suzuki(1986)]{SUZ1986}
H.~Suzuki.
\newblock Vibration and sound radiation of a piano soundboard.
\newblock \emph{Journal of the Acoustical Society of America}, 80\penalty0
  (6):\penalty0 1573--1582, 1986.

\bibitem[Touzé et~al.(2002)Touzé, Thomas, and Chaigne]{TOU2002}
C.~Touzé, O.~Thomas, and A.~Chaigne.
\newblock {Asymmetric non-linear forced vibrations of free-edge circular
  plates. Part 1: Theory}.
\newblock \emph{Journal of Sound and Vibration}, 258\penalty0 (4):\penalty0
  649--676, 2002.

\end{thebibliography}

\end{document}